\documentclass{ws-ijmpa}

\begin{document}

\markboth{J. Alfaro, A.A. Andrianov, L. Balart, D. Espriu}
{Hadronic string,  conformal invariance and  chiral symmetry}


\title{HADRONIC STRING, CONFORMAL INVARIANCE\\ 
AND CHIRAL SYMMETRY}

\author{\footnotesize J. ALFARO\footnote{jalfaro@puc.cl}}

\address{Facultad de F\'\i sica,
Universidad Cat\'olica de Chile\\
Vicu\~na Mackenna 486, Macul,   
Santiago de Chile, Casilla 306, Chile}

\author{\footnotesize A. A. ANDRIANOV\footnote{andrianov@bo.infn.it}}

\address{V.A. Fock Department of Theoretical Physics, St. Petersburg State
University,\\ 
1,ul. Ulianovskaya,
St. Petersburg 198504, Russia\\
and\\
Istituto Nazionale di Fisica Nucleare, Sezione di Bologna\\
via Irnerio 46,
Bologna 40126, Italy}

\author{\footnotesize L. BALART\footnote{lbalart@puc.cl}}

\address{Facultad de F\'\i sica,
Universidad Cat\'olica de Chile\\Vicu\~na Mackenna 486, Macul,   
Santiago de Chile, Casilla 306, Chile}

\author{D. ESPRIU\footnote{espriu@ecm.ub.es}}

\address{Departament d'Estructura i Constituents de la Mat\`eria,
Universitat de Barcelona\\
Av. Diagonal 647,  
Barcelona 08028, Spain}

\maketitle


\begin{abstract}
While it is clear that in some kinematic regime QCD can be described
by an effective (as opposed to fundamental) string theory,
it is not at all clear how this string theory should be. The `natural'
candidate, the bosonic string,  leads to amplitudes with 
the usual problems related to the existence of the
tachyon, absence of the adequate Adler zero,  and massless vector particles, 
not to mention the conformal anomaly.  The supersymmetric 
version does not really
solve most of these problems.  For a long time it has been believed 
that the solution 
of at least some of these difficulties is associated to a proper identification
of the vacuum, but this program has remained elusive. We show 
in this work how the first 
three problems can be avoided, by using
a sigma model approach where excitations above the correct
(chirally non-invariant) QCD vacuum are identified.  
At the leading order in a derivative
expansion we recover  the non-linear sigma model of pion interactions. At the
next-to-leading order the $O(p^4)$ Lagrangian of Gasser and Leutwyler is
obtained, with values for the coefficients that match the observed values.
We also discuss some issues related to the conformal anomaly.
\end{abstract}

\section{Introduction}

It seems that a paper with this title should necessarily start with a list
of the reasons that make us believe that there should be an 
effective description
of QCD in terms of strings\cite{string}. 

The more commonly cited
arguments are the dominance of planar diagrams
in the large $N$ limit\cite{largeN} `filling
in' a surface (interpreted as the world-sheet of a string),
the expansion in terms of surfaces built out of plaquettes
in strong-coupling lattice QCD\cite{lattice}, and the success of
Regge phenomenology\cite{regge}, which can ultimately be understood
in terms of string theory ideas (although, as we will discuss a little
bit later, the actual Regge theory that corresponds to QCD cannot
be derived, at present, from any known string theory).

To these we could add two more reasons. One is the appearance
in string theory of the universal (at long
distances) L\"uscher term\cite{ML}. The static inter-quark
potential provided by the string $V(r)=\sigma r + c$ gets modified
by quantum fluctuations by a Coulomb-like piece
$-\pi/12 r$. While for the heavy
quark bound state calculations it is perturbative QCD 
(complemented with some non-perturbative corrections) that is applicable,
and not the string regime, the L\"uscher term is quite useful.
Finally, and in a completely different context, namely that of
deep inelastic scattering, the evolution of the parton distribution
down to low values of $Q^2$ (around (2 GeV)$^2$)
leads\cite{evol} to a low $x$ behavior for the structure functions of the
form $x^{-1.17}$,
while Regge theory predicts $x^{-1}$, in striking good agreement.

In the last years the possibility of describing some non-abelian gauge
theories in terms of string theories has become a reality thanks
to the AdS/CFT relationship\cite{Mald}. Unfortunately, the gauge theories that
one can describe, or rather solve, in this way possess an unrealistically 
large number of supersymmetries. Breaking this large 
amount of supersymmetry to 
the realistic case
 of $N=0$ while decoupling the unwanted degrees of freedom from long-distance 
physics has proven to be a formidable task and one in which not much 
success has been attained yet.

It should be clear to the reader from the very beginning 
that we are not addressing 
in this paper these deep and fundamental issues. Yet, it is clear
that the mentioned theoretical developments give credence to the  idea
that there must be some kinematic regime where the simplest 
bosonic string, obtained after integrating out all the remaining degrees
of freedom made heavy by the breaking of supersymmetry, must provide 
an approximately valid description. 

In fact it is highly questionable
where one should expect a string description to be exactly valid for 
real, non-supersymmetric, four-dimensional, asymptotically free QCD, as    
a string does not seem to be the
natural language to understand high-energy processes in deep-inelastic 
scattering where the point-like structure of quarks and gluons is apparent.
We should probably be less ambitious and satisfy ourselves with
an effective description.
We subscribe this point of view and think of strings as
effective theories and not worry at all about their
mathematical consistency as fundamental objects.

It is surprising that even with this limited scope
all known string theories fail to 
provide a description of low energy QCD.
To see how this comes about let us recall
the original Veneziano amplitude\cite{Ven}. 
The original motivation for this {\it ad hoc} formula was to provide a
description of hadrons that manifestly exhibited duality. While the 
evidence for duality was rather weak\cite{harari}, at the 
time this amplitude sparked Nambu\cite{NG} to suggest 
that it should originate in bosonic string theory. 

After decorating the Veneziano amplitude 
with the appropriate Chan-Paton\cite{CP} factors\footnote{Due to
difficulties with unitarity only orthogonal groups or certain representations of unitary groups
can be introduced in this
way.}, it is supposed to describe the scattering amplitude of four
pions 
\begin{eqnarray}
{\cal A}(\pi^a \pi^b\to \pi^c \pi^d) \sim {\rm Tr}(T^a T^b T^c T^d) A(s,t)
+ {\rm non~cyclic~permutations},
\end{eqnarray}
\begin{eqnarray}
A(s,t)=
\frac{\Gamma(-\alpha(s))\Gamma(-\alpha(t))}{\Gamma(-\alpha(s)-\alpha(t))},
\end{eqnarray}
where $\alpha(s)=1+\alpha^\prime s$ is the Regge trajectory.
The Regge trajectory slope is determined, after a fit, to be
$\alpha^\prime \simeq 0.9 $ GeV$^{-2}$ \cite{lia} 
We immediately recognize that there are poles
in the $s$-channel whenever
$\alpha^\prime s= n-1$. Thus a tachyon is
present for $n=0$ and we see the first of the problems alluded to in
the abstract appearing. Furthermore the second ($J=1$) state (which we should
identify with the rho particle) is massless. This is the second 
one of the problems we alluded to. 

Shortly afterward, supersymmetry was introduced in string theory\cite{LS}, but
unfortunately most of the difficulties with the bosonic 
string are still present 
in supersymmetric string theory when one applies it to the
present context.
The relevant amplitude for our purposes is the four tachyon amplitude
\footnote{In supersymmetric theories one
usually performs the
GSO projection\cite{scherk}, projecting out the tachyon. However one may
choose not to do so and compute the four-tachyon amplitude,
supposed to describe scattering of spinless particles in the 
bosonic (Neveu-Schwarz) sector.}
\begin{eqnarray}
A(s,t)= \frac{\Gamma(1-\alpha_\rho(s))\Gamma(1-\alpha_\rho(t))}{
\Gamma(1-\alpha_\rho(s)-\alpha_\rho(t))}.
\label{neveu}
\end{eqnarray}
Now $\alpha_\rho(s)= 1 +\alpha^\prime s$ is the so-called $\rho$-trajectory
which in Regge theory describes exchanges of particles 
with positive $G$-parity.
This is the Lovelace-Shapiro amplitude and it contains no tachyonic poles.
Could it then be a candidate to describe
pion scattering? The answer is no. It does not have the appropriate Adler
zero, i.e. the property that at $s=t=0$ the pion scattering amplitude vanishes.
And although it has no tachyons in the intermediate channel, unfortunately,
the vertex which is supposed to represent the pion is a tachyon itself.
Furthermore, the first resonance in the 
$\rho$-trajectory (supposed to represent
the physical $\rho$ particle ) has zero mass. All in all, one is at the same
dead end.

A fix\cite{lovelace} to this problem consists in arbitrarily changing the 
intercept in $\alpha_\rho(s)$. Indeed, if we write $\alpha_\rho(s)
=\frac{1}{2} + \alpha^\prime s$ and use Eq.~(\ref{neveu}), the resulting
amplitude  has poles 
in the $s$ or $t$-channel when $\alpha^\prime s= n +\frac{1}{2}$.
It has no tachyons and the first pole, which is massive, can be
identified with the $\rho$ particle.
Furthermore, the previous amplitude has the right Adler zero. Expanding 
this amplitude in powers of $s$ and $t$ and comparing 
with the amplitudes obtained
from an effective Lagrangian, Polyakov and Vereshagin\cite{PV}
found that
$L_1=\frac{1}{2} L_2,\qquad L_2=\frac{F_\pi^2}{8m_\rho^2}\ln 2, \qquad
L_3=-2L_2 $, which numerically turn out to be quite acceptable
values\footnote{The relation $L_1 = 1/2 L_2 = -1/4 L_3$ was
established earlier
in bosonization models \cite{AnBo}
and in the chiral quark model\cite{Espr} by means of
a derivative expansion of quark determinant. However at that time its possible
connection with a string 
description of QCD was not recognized.}. Unfortunately, no known
string 
theory leads to such a fix in the
intercept\footnote{It appears possible to modify to some extent the Regge 
trajectory for low values of $J, M^2$, and hence the intercept 
by adding additional
terms in the string action, rigidity, for instance\cite{rigid,Curtright}
or imposing constraints on the string motions\cite{pronko}.
However spectroscopy suggest that while such terms might exist
they are probably of limited influence. Moreover
the conformal invariance of a modified string becomes questionable.}.

It is believed that the ultimate reason for the
presence of a tachyon in the spectrum lies in a wrong choice
of the vacuum\cite{tachyon}. Since the choice of the spin zero vertex operator,
$V(k)=:\exp{ikx}:$, is based on the Lorentz properties alone, it is the same
both for scalar and pseudoscalars and, accordingly, both scalars and
pseudoscalars have tachyonic poles in the $s$-channel on account of
parity conservation. The situation is thus parallel
to the one in multicomponent $\phi^4$ theory with $V(\phi)=-\mu^2 +
\lambda \phi^4$, where
perturbing around $\phi=0$ gives negative $m^2$ values
for all components. It is natural
to assume that the amplitudes obtained through the use
of the canonical vertex operators correspond to
(unphysical)
amplitudes for excitations perturbed around the
wrong, unphysical vacuum. These ideas have in fact been around 
for a long time, but no one appears to have implemented them
in any practical way.

In a previous paper\cite{ADE} by some of the present authors
a possible line of attack was suggested. The idea is the 
following. Given that with the present string theory technology
it is impossible to find the `true' vacuum, let us assume
that the vacuum is non-perturbative in string creation
and annihilation operators and that it actually corresponds
to the true QCD vacuum.
In this case, the
relevant (lightest) degrees of freedom are the
ones emerging after the spontaneous breaking of
chiral symmetry. In the physical vacuum of QCD there
is a clear distinction between scalars (sigma particle) and pseudoscalars
(pions). 
The sigma particle is just 
another hadronic resonance with a mass in the 1 GeV region
and without any specially important role to play.
The massless pseudoscalars, on the other hand,  can be collected in a
unitary matrix $U(x)$ which under chiral transformations
belonging to $SU(3)_L\times SU(3)_R$
transforms as
$U(x)\to U^\prime(x) = L U(x) R^\dagger$ and describes excitations
around the non-perturbative vacuum. From the string point of view
$U(x)$ is nothing but a
bunch of couplings involving the string variable $x$. Our goal is to find 
a consistent string propagation in this non-perturbative background.

A crucial property of string theory is, undoubtedly, conformal invariance. 
This amounts to demanding that the theory
is independent of the specific conformal factor chosen to describe
the two-dimensional world sheet. While this is
a desirable and, as a matter of fact, a crucial 
property of fundamental strings, it might not be
necessarily so for the QCD string (if we look at the QCD string with a magnifying glass
we shall eventually see quarks and gluons, not the string itself!) . However in the limit of large $N_c$
the QCD hadronic amplitudes are saturated by gluon fishnet diagrams among which any higher order
samples give, in principle, a comparable contribution. Thereby it
appears plausible that there should be no dependence on a specific
choice of configuration describing a dominating gluon propagation. Therefore the hadronic string action
should obey reparameterization invariance of diagram surfaces and
conformal invariance. 
Since conformal invariance must hold when 
perturbing the string around any vacuum, perturbative or not, we have
a powerful tool at our disposal, namely to couple the chiral field $U(x)$
to the string degrees of freedom and demand conformal invariance. Exactly
in the same way as Einstein equations are derived from
string theory\cite{callan} by requiring the vanishing of the beta function
for an external metric $G_{\mu\nu}(x)$.

This approach was followed in \cite{ADE} with mixed success. The lowest
order chiral Lagrangian, or rather the equations of motion coming from it, 
were obtained and the whole approach seemed consistent. Unfortunately,
the $O(p^4)$ coefficients were identically zero. Experimentally these
are numbers of order $10^{-3}$. We have now found out the reason for this
failure. It turns out that the action considered in \cite{ADE} is only
a particular case of the general case, and one that does not allow a
proper treatment of the unitarity condition of the external source $U(x)$. When
the general case is considered, unitarity and locality can be implemented
consistently ---at least at the order we have worked--- and, as a consequence
of the new terms in the action, new divergences and hence new contributions
to the beta function for $U(x)$ appear. These lead to $O(p^4)$ terms
in the effective Lagrangian. These are the results we would 
like to report here.

\section{Coupling pions to the QCD string}

The hadronic string in the conformal gauge is described by 
the conformal field theory action in the four dimensional Euclidean space-time
\footnote{The direct evidences for dimensions of the
string world sheet\cite{fub,vent} (d=2) and of the 
target space\cite{cud} (D=4) were
found from the analysis of meson state densities at high energies
 as compared to the Hagedorn-like experimental growth of the latter ones.}
\begin{equation}
{\cal W}_{str}=\frac{1}{4\pi\alpha'}\int d^{2+\epsilon}\sigma
\left(\frac{\varphi}{\mu}\right)^{-\epsilon} 
\partial_i x_\mu \partial_i x_\mu, 
\label{string}
\end{equation} 
where for $\epsilon = 0$ one takes 
$$x_\mu = x_\mu(\tau, \sigma);\quad 
-\infty <\tau< \infty, 0< \sigma <\infty;\quad
i = \tau,\sigma \quad \mu=1,...,4.$$ 
The conformal factor $\varphi(\tau, \sigma)$ is 
introduced to restore the conformal 
invariance in $2+\epsilon$ dimensions, and this is 
the only way it enters the theory. The Regge trajectory slope
(related to the inverse string tension) is known
to be universal $\alpha' \simeq 0.9$ GeV$^{-2}$ \cite{lia}.

We would like to couple in a chiral invariant manner
the matrix in flavor space $U(x)$ containing  the meson fields to the
string degrees of freedom while preserving
general covariance in the two dimensional coordinates and conformal
invariance under local scale transformations
of the two-dimensional metric tensor.
The equations of motion for the $U(x)$ field will then be obtained from the
condition that the quantum theory must be
conformally invariant, i.e. the
$\beta$ functional for the $U(x)$ couplings must vanish.

Since the string variable $x$ does not contain any flavor dependence,
we have to invent a way to couple it to the background $U(x)$ variable.
We introduce two dimensionless Grassmann variables (`quarks'), or 
rather several families
of them, living on the
boundary of the string sheet: 
$\psi_L(\tau),\psi_R(\tau)$ which transform in the fundamental representation
of the light flavor group ($SU(2)$ in the present paper).
A local hermitean action $S_b = \int d\tau L_f$ is introduced on the boundary
$ \sigma =0$ to describe the interaction with background chiral fields 
$U(x(\tau)) = \exp(i \pi(x)/f_\pi)$ where the normalization scale $f_\pi\simeq 93 MeV$, 
the weak pion decay constant, is introduced to relate the field $\pi(x)$ to a $\pi$-meson one.

The boundary Lagrangian is chosen to be reparameterization invariant
and in its simplest minimal form reads 
\begin{eqnarray}
L_f&=&\frac12 i \left(\bar\psi_L U (1 - z) \dot\psi_R  - 
\dot{\bar\psi_L} U (1 +z)
\psi_R \right.\nonumber\\
&&\left.+ \bar\psi_R U^+ (1 + z^*)
\dot\psi_L - \dot{\bar\psi_R} U^+ (1 - z^*) \psi_L\right), \label{lag}
\end{eqnarray}
herein and further on a dot implies a $\tau$ derivative: 
$\dot\psi \equiv d\psi/d\tau$. In order to arrive to Eq.~(\ref{lag})
a number of field redefinitions have been made. It is impossible to simplify
Eq.~(\ref{lag}) any further. Details are given in Appendix A.

A further restriction is obtained by requiring $CP$ invariance.
There are two $CP$-like transformations. The first one is
\begin{equation}
U \leftrightarrow U^+, \quad \psi_L  \leftrightarrow  \psi_R.
\label{CP1}
\end{equation}
The above Lagrangian is $CP$ symmetric for $ z = - z^* = ia$.
The second one is
\begin{equation}
U \leftrightarrow U^+, \quad \psi_L  \rightarrow U^+ \psi_L, \quad
 \psi_R  \rightarrow U \psi_R.
\end{equation}
Under this transformation the Lagrangian becomes invariant only for
$z = 0$. We interpret the first $CP$ transformation as the physical one
and the one which one should require of a Lagrangian describing strong interactions.

The above coupling may appear surprising at first and somewhat {\it ad hoc}.
To see that this is not so, let us expand the non-linear field
$U(x)$, i.e. $U(x)\simeq 1 + i\pi(x)/f_\pi + ...$ and retain the first two
terms. The first term just gives rise to a $\theta$-function propagator
which eventually leads to the familiar ordering in the usual
string amplitudes $t_1< t_2< ....$. The second term just provides
(after integrating the fermions out) the usual (tachyonic!) vertex.
In short, if we ignore the non-linearities in the theory we
are back to the usual difficulties.

It is easy to see that the previous action is invariant under general
coordinate
transformations of the two dimensional world sheet.
The fermion action is automatically conformally invariant, because it does
not contain the two dimensional world sheet metric tensor since it can be
written as a line integral.

In \cite{ADE} the cases $z=\pm 1$ were considered. None of them
is a valid one on symmetry grounds.

\section{Diagrammar}

Now we expand $U(x(\tau))$ around a constant background $x_0$ and look
for the potentially divergent
one particle irreducible diagrams (1PI).
We classify them according to the number of loops. Each additional
loop comes with a power of $\alpha^\prime$.

We expand the function $U(x)$ in powers of the string coordinate 
field $x_\mu(\tau) =x_{0\mu} + \tilde x_\mu(\tau) $ around a constant $x_0$
which is the translational zero mode of the string
\begin{eqnarray}
U(x) &=& U(x_0) + \tilde x_\mu(\tau) \partial_\mu U(x_0) + 
\frac12 \tilde x_\mu(\tau) 
\tilde x_\nu(\tau) \partial_\mu\partial_\nu U(x_0) +\ldots \nonumber\\
&\equiv&  U(x_0) + {\cal V}(\tilde x). 
\label{expan}
\end{eqnarray}
One can find a resemblance to the familiar 
derivative expansion of chiral perturbation theory. Indeed
perturbation theory in the operators (\ref{expan}) makes sense as a low 
momentum expansion 
which is presumably valid up to momenta
approaching to the mass of the first massive resonance ($\rho$ meson
etc.). In the present case $\alpha^\prime$ is the dimensional 
parameter normalizing the above expansion.

The free fermion propagator is
\begin{equation}
\langle\psi_R (\tau) \bar\psi_L(\tau')\rangle = U^{-1} (x_0) \theta(\tau - \tau').
\end{equation}
If we impose $CP$ symmetry then
\begin{equation}
\langle\psi_L (\tau) \bar\psi_R(\tau')\rangle= 
\langle\psi_R (\tau) \bar\psi_L(\tau')\rangle^\dagger = U (x_0) \theta(\tau - \tau'),
\end{equation}
for unitary chiral fields $U(x)$.

The free boson propagator projected on the boundary is
\begin{equation}
\langle x_\mu(\tau) x_\nu(\tau')\rangle = \delta_{\mu\nu}
\Delta (\tau -\tau'),\quad  
\Delta (\tau \rightarrow \tau') = \Delta (0) \sim
\frac{\alpha'}{\epsilon}, \quad \partial_\tau\Delta (\tau \rightarrow \tau') = 0,
\end{equation}
the latter results hold in dimensional regularization (see below).

In order to make contact between dimensional regularization, 
a short-distance
cut-off (which we shall later use) and Regge phenomenology
we need to unambiguously fix the
normalization of the string propagator. This can be inferred from the  
definition of the kernel of the N-point tachyon amplitude for the 
open string\cite{Rebbi}. The Veneziano 
amplitude corresponds to the insertion of vertex operators
$:\exp(i k^{(j)}_\mu x^\mu(\tau_j)):$ on the boundary of the
string. After resolution of the Gaussian integral one obtains
for the kernel of the generalized beta-function 
\begin{eqnarray}
\langle\prod_j :\exp(i k^{(j)}_\mu x_\mu(\tau)):\rangle &=&
\exp\left( -\frac12 \sum_{j\not=l} 
k^{(j)}_\mu k^{(l)}_\mu\Delta (\tau_j -\tau_l)\right)\nonumber\\
&\equiv&\prod_{j>l} |\tau_j -\tau_l|^{2 \alpha^\prime k^{(j)}k^{(l)}}, 
\label{norma}
\end{eqnarray}
which unambiguously prescribes
\begin{equation}
\Delta (\tau_j -\tau_l) = - 2\alpha' \ln(|\tau_j -\tau_l|\mu). \label{prop} 
\end{equation}
The $\mu$ dependence does not show up in (\ref{norma}) due to 
energy-momentum conservation.

Keeping in mind this definition let us determine the string 
propagator in dimensional regularization, restricted on the boundary. 
First we calculate the
momentum integral in $2+\epsilon$ dimensions
\begin{eqnarray}
\Delta_\epsilon (\tau) &=& 4\pi\alpha' 
\left(\frac{\varphi}{\mu}\right)^{\epsilon}
 \int \frac{d^{2+\epsilon}k}{(2\pi)^{2+\epsilon}}
\frac{\exp(ik_0\tau)}{k^2}\nonumber\\
 &=& \alpha' \Gamma\left(\frac{\epsilon}{2}\right) \ 
\left|\frac{\tau\mu\sqrt{\pi}}{\varphi}\right|^{-\epsilon}\nonumber\\
&\stackrel{\epsilon \to 0}{=}&  2\alpha'\left[\frac{1}{\epsilon} + C - 
\ln\left(\frac{\tau\mu}{\varphi}\right)\right] + {\cal O}(\epsilon).
\label{epsprop}
\end{eqnarray}
A dimensionally regularized propagator properly 
normalized to reproduce (\ref{prop})  can be constructed  
by subtracting from (\ref{epsprop}) its value at $\tau\mu =1$
where (\ref{prop}) should vanish
\begin{eqnarray}
\Delta_\epsilon (\tau)|_{reg} &=& \alpha'
\Gamma\left(\frac{\epsilon}{2}\right)
\  \left\{
\left|\frac{\tau\mu\sqrt{\pi}}{\varphi}\right|^{-\epsilon}
- \left|\frac{\sqrt{\pi}}{\varphi}\right|^{-\epsilon}\right\} \nonumber\\
&\stackrel{\epsilon \to 0}{=}& 
- 2\alpha' \ln|\tau\mu|+ {\cal O}(\epsilon). \label{reg}
\end{eqnarray}
Therefrom one finds unambiguously the relation
\begin{equation}
\Delta (0) =  - \alpha'
\Gamma\left(\frac{\epsilon}{2}\right) 
\left|\frac{\sqrt{\pi}}{\varphi}\right|^{-\epsilon}
\stackrel{\epsilon \to 0}{=}
- 2\alpha'\left[\frac{1}{\epsilon} + C + 
\ln\varphi\right] + {\cal O}(\epsilon) \equiv \Delta_\epsilon 
- 2\alpha'\ln\varphi, \label{Dzero}
\end{equation}
where in the spirit of dimensional regularization
 we have assumed that $\epsilon < 0$ and hence
the first term in (\ref{reg}) vanishes at $\tau =0$.

The two-fermion, $N$-boson vertex operators are generated by the expansion 
(\ref{expan}) and they appear with an extra sign $(-) = i^2$ 
following the definition for the generating functional 
$ Z_b = \langle\exp(i S_b)\rangle$ and Eq.~(\ref{lag}). In particular, for the $L \to R$ transition one has
\begin{equation}
V = - \frac12 \left((1-z){\cal V}(\tilde x) \partial_\tau  + (1+z)\partial_\tau \left[{\cal V}(\tilde x)\ldots\right]\right),
\end{equation}
and for the  $R \to L$ transition the hermitean 
conjugated vertex $V^+$ appears.
The corresponding Feynman rule for the 2-fermion, N-boson vertex of the $L \to R$ transition is
\begin{eqnarray}
{\cal V}_N = - \frac{1}{2n!} \partial_A \delta(A-B)& \partial_{\mu_1}\ldots \partial_{\mu_N} U(x_0)
&\left[(1-z)\delta(A-\tau_1)\cdots\delta(A-\tau_N) +\right.\nonumber\\
&&\left.+(1+z)\delta(\tau_1 -B)\cdots\delta(\tau_N -B)\right], \label{Nvert}  
\end{eqnarray}
where $A,B$ are proper-time values for the left- and right-handed fermions and $\tau_1, \ldots,\tau_N $
are proper-time values for the boson field-string variables.

From this point on it is quite straightforward to proceed with the
renormalization process. We shall determine the counterterms required
to make the beta functional for the coupling $U(x)$ vanish up to
the two loop level. In spite of the relative complexity of the Feynman rules,
the fact that we are working with a boundary field theory is crucial in making
the calculation manageable. In fact most diagrams can be
determined by simply playing with integration by parts and using
basic properties of the Dirac delta function. Yet the renormalization
is quite non-trivial and the ultraviolet structure of the counterterms is 
surprisingly quite complex. It is thanks to this complexity that non-zero 
values for the $O(p^4)$ coefficients can be obtained. In fact we believe
that some of the results presented in this work can have some bearing on
more general discussions involving fundamental strings too.

In what follows we retain not only the singular parts of one-loop
diagrams  but also the finite ones 
as they will be necessary to construct two-loop diagrams.

\section{Renormalization of the fermion propagator at one loop}

To avoid clutter the main body of the paper we have relegated
the detailed derivation of the different Feynman diagrams to the appendixes.
Since the present calculation is somewhat non-standard, we provide
the technical details there.

Using the set of Feynman rules described in the previous section one
arrives to the following result for 
the divergent part of the propagator (Appendix B),
\begin{equation}
\theta(A - B)\frac12 \Delta (0) U^{-1} 
\left\{- \partial^2_{\mu} U + 
\frac{3 + z^2}{2}\partial_\mu U U^{-1}\partial_\mu U\right\} U^{-1}.
\end{equation}
This divergence is eliminated by introducing 
an appropriate counterterm $U \to U+\delta U$
\begin{equation}
\delta U = \Delta(0) \left[\frac12 \partial^2_{\mu} U - 
\frac{3 + z^2}{4}\partial_\mu U U^{-1}\partial_\mu U\right]= 0. \label{resym}
\end{equation}
Conformal symmetry is restored (the beta-function is zero) if
the above contribution vanishes.

Let us find out for which value of $z$ this variation of $U$ is 
compatible with 
its unitarity.
\begin{equation}
\delta (U U^+)= U \cdot\delta U^+ + \delta U\cdot U^+ = 0. \label{unit}
\end{equation}
A simple calculation shows that this
takes place for $z = \pm i$. 
For other values of $z$ eq.(\ref{resym}) entails
$\partial_\mu U U^{-1}\partial_\mu U = 0$ which has only a trivial constant solution in
Euclidean space-time. In \cite{ADE} this value of $z$ was not
considered and thus unitarity of $U$ was not properly taken
into account.

When $z = \pm i$ and before the unitarity constraints are imposed the local classical action which 
has (\ref{resym}) as equation of motion is
\begin{equation}
W^{(2)} = \frac{f_\pi^2}{8}\int d^4 x \mbox{\rm tr}\left[\partial_\mu U
\partial_\mu U^{-1} + \partial_\mu U^+
\partial_\mu (U^+)^{-1}\right]. \label{weinb}
\end{equation}
For other values of $z \not= \pm 1; \pm i$ the related local action 
is unknown. The above Lagrangian is of course the well known non-linear 
sigma model which is commonly employed to describe pion interactions.

We have thus succeeded in finding the action induced by the QCD string.
It has all the required properties of locality, chiral symmetry 
and proper low momentum behavior (Adler zero). Furthermore, it describes
massless pions. $f_\pi$, the overall normalization scale, cannot be
predicted from these arguments.

To this point we have been quite successful in our program, but of course
no real predictability has been achieved yet. Indeed we knew the form of this
action from general principles, even though it is nice to see that 
things work out consistently. We have to turn to the 
$O(p^4)$ effective Lagrangian to get non-universal results.  

\section{Renormalization of the vertices at one loop}

In order to proceed to a two loop calculation we shall need in addition
to the counterterms for the one-loop propagator (which we just got)
the counterterms for the vertex with two fermion lines and one and
two boson lines, $x^\mu$, respectively. We also have to check whether 
the minimal
Lagrangian (\ref{lag}) is sufficient to renormalize also the vertices. 
It turns out that, 
in fact, it is not.

  Let us obtain the divergences for vertices with external boson lines.
We introduce an external background boson field $\bar x_\mu $ 
to describe vertices with
several boson legs and split $ x_\mu =  \bar x_\mu + \eta_\mu $. 
The free propagator 
for the fluctuation field $\eta_\mu $ coincides with the one for  $x_\mu $.

Then the  total divergence in the vertex with two fermions and one boson line
is (see Appendix C)
\begin{eqnarray}
&& \theta(A - B)\frac14 \Delta (0) U^{-1}\left\{\bar x_\mu(A)
(1+z)\left[- \partial_{\mu}(\partial^2 U) + 
2 \partial_\nu U U^{-1}\partial_\mu\partial_\nu U \right.\right.\nonumber\\
&&\left.\left.+ (1+z)\partial_\mu \partial_\nu U U^{-1}\partial_\nu U 
- 
\frac12 (1+ z)(3 - z) 
\partial_\nu U U^{-1} \partial_\mu U U^{-1}\partial_\nu U\right] \right.\nonumber\\
&&\left. +\bar x_\mu(B)
(1-z)\left[- \partial_{\mu}(\partial^2 U) + 
(1-z) \partial_\nu U U^{-1}\partial_\mu\partial_\nu U +
2\partial_\mu \partial_\nu U U^{-1}\partial_\nu U\right.\right.\nonumber\\
&&\left.\left. - 
\frac12 (1- z)(3 + z) 
\partial_\nu U U^{-1} \partial_\mu U U^{-1}
\partial_\nu U \right]\right\} U^{-1} \nonumber\\
&& \equiv  - \frac12 \theta(A - B) U^{-1} \left[\bar x_\mu(B) 
\Phi^{(1)}_\mu + \bar x_\mu(A) \Phi^{(2)}_\mu 
\right]U^{-1}. \label{div1}
\end{eqnarray}
Let us now amputate the 
fermion legs and evaluate the pure
vertex divergences. The amputation rules are
\begin{eqnarray}
&&\bar x_\mu(A) \theta(A - B) = - \int d\tau  \partial_\tau
\theta(A - \tau) \bar x_\mu(\tau) \theta(\tau - B),\nonumber\\
&& \bar x_\mu(B) \theta(A - B) = \int d\tau  
\theta(A - \tau) \bar x_\mu(\tau) \partial_\tau \theta(\tau - B).
\end{eqnarray}
Then the divergent part (\ref{div1}) can be reproduced by the following
operator in the Lagrangian
\begin{equation}
\frac{i}{2} \left( \bar\psi_L \Phi^{(1)} \dot\psi_R  -\dot{ \bar\psi_L} 
\Phi^{(2)}
\psi_R\right) + \mbox{\rm h.c.},\qquad 
\Phi^{(1,2)} \equiv \bar x_\mu(\tau)\Phi^{(1,2)}_\mu,
\end{equation}
where the vertex matrices $\Phi^{(1,2)}_\mu$ can be rearranged as follows
\begin{eqnarray}
\Phi^{(1)}_\mu &=&  \Delta (0)\left\{ (1-z)  \partial_{\mu}\left( 
\frac12 \partial^2 U
- \frac{3 + z^2}{4}  \partial_\nu U U^{-1}\partial_\nu U \right)\right.\nonumber\\
&&\left. - \frac{1 - z^2}{2}\left( 
\frac{1 - z}{2} \partial_\mu \partial_\nu U U^{-1}\partial_\nu U 
 - \frac{1 + z}{2} \partial_\nu U U^{-1}\partial_\mu\partial_\nu U \right.
\right.\nonumber\\
&&\left.\left.+ z \partial_\nu U U^{-1} \partial_\mu U U^{-1} \partial_\nu U 
\right)\right\}\nonumber\\
&\equiv&  (1-z)  \partial_{\mu}\left(\delta U\right)  - \phi _\mu\nonumber\\
\Phi^{(2)}_\mu &=&  (1+ z)  \partial_{\mu}\left(\delta U\right)  + \phi _\mu .\label{adddiv}
\end{eqnarray}
The terms proportional to derivatives of $\delta U$ 
are automatically eliminated by the redefinition of $U$ that
one performs to renormalize the one-loop propagator (and it of course
vanishes if the equations of motion are imposed). 
But the part proportional to $\phi _\mu$ remains and to absorb 
these divergences new counterterms are required.
Evidently, these terms come out of the following terms in the Lagrangian:
\begin{eqnarray}
\Delta L_{div.} &=& \frac{i}{4}\Delta (0) (1 - z^2)\bar\psi_L\left( 
\frac{1 - z}{2} \partial_\nu \dot{U} U^{-1}\partial_\nu U 
 - \frac{1 + z}{2} \partial_\nu U U^{-1}\partial_\nu \dot{U} \right.\nonumber\\
&&\left.
+ z \partial_\nu U U^{-1} \dot{U} U^{-1} \partial_\nu U \right)\psi_R  
+ \mbox{\rm h.c.} \label{newver}
\end{eqnarray}
Therefore the counterterms 
required to eliminate
the additional divergences for the vertex with one boson and two
fermion lines can be parameterized with three bare 
constants $g_1$ , $g_2$ and $g_3$, which are real 
if the $CP$ symmetry for $z = - z^*$ holds
\begin{eqnarray}
\Delta L_{bare}&= &\frac{i}{8} (1 - z^2)\bar\psi_L\left( 
(g_1 - z g_2) \partial_\nu \dot{U} U^{-1}\partial_\nu U 
- (g_1 + z g_2) \partial_\nu U U^{-1}\partial_\nu \dot{U} \right.\nonumber\\
&&\left.+  2z g_3 \partial_\nu U U^{-1} \dot{U} U^{-1} \partial_\nu U 
\right)\psi_R  + \mbox{\rm h.c.} \label{count}
\end{eqnarray} 
Renormalization is accomplished by redefining the couplings $g_i$
in the following way
\begin{equation}
g_i = g_{i,r} - \Delta(0). \label{gren}
\end{equation}
The constants $g_{i,r}$ are finite, but in principle scheme dependent, and 
from (\ref{Dzero}) it follows that a logarithmic dependence of the
bare couplings
on the conformal factor $\varphi$ is introduced along the renormalization
process.
The counterterms are of  higher dimensionality than 
the original Lagrangian (\ref{lag}) and therefore the couplings $g_i$ are 
of dimension $M^{-2}$. Since (\ref{lag}) was actually the most 
general coupling permitted by the symmetries of the model, in 
particular conformal invariance, one is lead to the conclusion that
conformal symmetry is broken, already at tree level,  
by these couplings, unless they 
happen to vanish. Since they are dimensional, it is natural to
normalize them by the only dimensional parameter, namely
$\alpha^\prime$.

Even if the new couplings are dimensional, it turns out that at the order
we are computing, the trace of the energy-momentum tensor 
is still vanishing once the requirements of unitarity of $U$ are 
taken into account (see Appendix D) and therefore conformal invariance is not broken 
at the order
we are working. At higher orders in the $\alpha^\prime$ expansion
further counterterms may be however required in order 
to ensure conformal invariance
perturbatively. We postpone a more detailed discussion to the final
sections.

One can introduce the running ``effective" couplings
\begin{equation}
g_i  + \Delta_\epsilon = g_{i,r} +2\alpha'\ln\varphi \equiv  g_{i}^\varphi,
\end{equation}
One-loop conformal  invariants are $g_1^\varphi - g_2^\varphi$ and 
$g_1^\varphi - g_3^\varphi$. 
At any rate, the dependence of the new couplings on the Liouville mode
is determined. 

In any case,
the appearance of new vertices changes the  
fermion propagator due to the diagrams discussed in
Appendix E. One obtains from such terms (which are of higher 
order in derivatives) the following contribution to the propagator 
\begin{eqnarray}
&&\theta(A-B)\frac{1}{16}\Delta(0) (1-z^2)
U^{-1} \left\{2 (g_{1,r} - z^2 g_{2,r})
 \partial_{\rho} U  U^{-1} \partial_\mu\partial_\rho U  U^{-1}\partial_{\mu} U\right.\nonumber\\ 
&&\left.- (1+z) (g_{1,r} + z g_{2,r}) \partial_{\rho} U  U^{-1} \partial_\mu U  U^{-1} \partial_\rho\partial_{\mu} U\right.\nonumber\\ 
&&\left.
- (1-z) (g_{1,r} - z g_{2,r}) \partial_{\rho}\partial_\mu U  U^{-1} \partial_\rho U  U^{-1}\partial_{\mu} U\right.\nonumber\\ 
&&\left.
+4z^2 g_{3,r} \partial_{\rho} U  U^{-1} \partial_\mu U  U^{-1} \partial_\rho U  U^{-1}\partial_{\mu} U
\right\} U^{-1} \nonumber\\
&\equiv& - \theta(A-B)\Delta(0) U^{-1}  \delta^{(4)}U  U^{-1} \label{dren}
\end{eqnarray}
we shall denote this contribution by $\theta(A-B) d_g$.
One should  add 
the divergence contained in $d_g$ to the one-loop result, thereby 
modifying the 
$U$ field renormalization and equations of motion
\begin{equation}
\bar\delta U = \Delta(0) \left[\frac12 \partial^2_{\mu} U - 
\frac{3 + z^2}{4}\partial_\mu U U^{-1}\partial_\mu U + \delta^{(4)}U\right]= 0.
\end{equation}
This is, in fact, the source of the much sought after $O(p^4)$ terms.

We note that this new
contribution (\ref{dren}) is proportional to $1-z^2$ and it was thus absent
in \cite{ADE}.
At this point we have to ask whether
such equation of motion may be derived from a local effective Lagrangian
containing both  dimension 2 and dimension 4  operators. This would
then constitute the effective Lagrangian derived from the string model. 
However at this point this question is too premature to formulate.
Two-loop diagrams generated from 
(\ref{lag}) could certainly produce similar contributions and, as a matter of 
fact, so could new counterterms from diagrams with two fermions and two
boson lines, should they require an additional counterterm. In fact
it can be seen that the above equation of motion cannot be derived
from a local Lagrangian involving the unitary matrix $U$. The
requirements of locality and unitarity would force $g_{j,r} = 0$, 
so this should not be the full answer. 

Before concluding this section and moving to the other contributions
we have just mentioned,
we calculate the renormalization of the vertex with two fermion
and two boson lines, as this is also required as a counterterm.
Let us summarize the divergent structure for these diagrams (see Appendix F)
\begin{eqnarray}
&&  \theta(A - B)\frac14 U^{-1}\left\{\bar x_\mu(A) \bar x_\nu(A)\left[
 \partial_\mu\partial_\nu (-\delta U) (1+z) - \phi_{\mu\nu}\right]\right.\nonumber\\
&&\left. +  \bar x_\mu(B) \bar x_\nu(B)\left[
 \partial_\mu\partial_\nu (-\delta U) (1-z) + \phi_{\mu\nu}\right] \right.\nonumber\\
&&\left. + \Delta (0)frac{1-z^2}{2}\int^A_B d\tau\left[ - 
\bar x_\mu(\tau) \dot{\bar x}_\nu(\tau)
 \partial_{\rho} U  U^{-1} \partial_\mu U  U^{-1} 
\partial_\nu\partial_{\rho} U (1-z)\right.\right.\nonumber\\
&&\left.\left. 
+ \dot{\bar x}_\mu(\tau) \bar x_\nu(\tau)  
\partial_{\rho}\partial_\mu U  U^{-1} \partial_\nu U  
U^{-1}\partial_{\rho} U (1+z)\right.\right.\nonumber\\
&&\left.\left. +(\bar x_\mu(\tau) 
\dot{\bar x}_\nu(\tau) -\dot{\bar x}_\mu(\tau) \bar x_\nu(\tau))
 \partial_{\rho}\partial_\mu U  
U^{-1} \partial_\nu\partial_{\rho} U\right]\right\} U^{-1}, \label{secvar}
\end{eqnarray}
where
\begin{eqnarray}
\phi_{\mu\nu}&=& \Delta (0)\frac{1-z^2}{2} \left[ \frac{1-z}{2}
\partial_{\rho}\partial_\mu \partial_\nu U  U^{-1}\partial_{\rho} U
- \frac{1+z}{2} \partial_{\rho} U  U^{-1} \partial_\mu\partial_\nu\partial_{\rho} U\right.\nonumber\\
&&\left.- z  \partial_{\rho}\partial_\mu U  U^{-1} 
\partial_\nu\partial_{\rho} U +  
z \partial_{\rho} U  U^{-1} \partial_\mu\partial_\nu U  
U^{-1}\partial_{\rho} U\right.\nonumber\\
&&\left.+2z( \partial_{\rho} U  U^{-1} \partial_\mu U  U^{-1} \partial_\nu\partial_{\rho} U +
\partial_{\rho}\partial_\mu U  U^{-1} \partial_\nu U  U^{-1}\partial_{\rho} U)\right.\nonumber\\
&&\left. - 2z \partial_{\rho} U  U^{-1} \partial_\mu U  U^{-1} 
\partial_\nu U  U^{-1}\partial_{\rho} U\right]
\end{eqnarray}
One can check that the terms encoded in $\phi_{\mu\nu}$, together with the last
contribution in Eq.~(\ref{secvar}) combine precisely as a
second variation of the additional interaction vertices (\ref{newver}) 
in coordinate fields. Therefore their renormalization is completely performed 
with the help of counterterms (\ref{count}) and no additional counterterms
or operators appear.

It can also be seen that, in fact,
 any diagram with an arbitrary number of external
boson lines and two fermion lines, i.e. any vertex of those generated by the
perturbative expansion of
(\ref{lag}) is rendered finite by the previous counterterms. This completes
the renormalization program at one loop.

\section{The fermion propagator at two loops}

The
two-loop contributions to the fermion propagator can be obtained from
one-loop diagrams with two external boson legs by joining the latter
with a boson propagator. A factor $1/2$ has to be added.
One must include not 
only one-particle irreducible diagrams but also
one-particle reducible ones and consider  
both divergent and finite parts.
 
There are 10 two-loop diagrams which are listed in Appendix G.
The divergences in the propagator at two-loops can be presented 
separated into five pieces
\begin{equation}
\theta(A-B)[ d_{I} + d_{II}+ d_{III}+ d_{IV}+ d_{V} ].
\end{equation}
The first and second piece contain the double divergence $ \Delta^2(0)$, 
the third, fourth and fifth pieces reveal only a single divergence  
$ \Delta (0)$. Finite parts are irrelevant for the present 
discussion.

\begin{table}[h]
\tbl{Chiral field structures proportional to the double
divergence $\Delta^2 (0)$ appearing in the two loop contribution
 to the fermion propagator.}
{\begin{tabular}{@{}cccc@{}} \toprule
{\bf CF structure}& $d_I$& $d_{II}$ &
{\bf  Total} \\  \colrule
$\mu^2\rho^2$& $-\frac18$& 0 & $-\frac18$  \\ 
$\mu^2\rho--\rho$& $\frac{3+z^2}{8}$ &$ 0$ &$ \frac{3+z^2}{8}$   \\  
$\rho--\mu^2\rho$& $ \frac{3+z^2}{8}$& $0$ & $\frac{3+z^2}{8}$   \\  
$\rho--\mu^2--\rho$&$ -\frac{3+z^2}{8} $&$0 $&$ - \frac{3+z^2}{8}$  
\\  
$\rho--\mu--\mu\rho $&$ -\frac{3+z^2}{8}  -\frac{(3+z^2)^2}{32} 
$&$  -\frac{(1-z^2)(1+z)^2}{32} $&$-\frac{11+z+5z^2-z^3}{16}$   \\  
$\mu\rho--\mu--\rho$&$ -\frac{3+z^2}{8}  -\frac{(3+z^2)^2}{32}
 $&$-\frac{(1-z^2)(1-z)^2}{32} $&$ -\frac{11-z+5z^2+z^3}{16}$ \\ 
$\mu\rho--\mu\rho$&$  \frac{3+z^2}{8} $&$ 0 $&$  \frac{3+z^2}{8}$
  \\  
$\rho--\mu--\mu--\rho$&$ \frac{3+z^2}{8}  +\frac{(3+z^2)^2}{32} $&$ 0 $&$ 
 \frac{(3+z^2)(7+z^2)}{32}$ \\  
$\mu--\mu\rho--\rho $ &$  -\frac{(3+z^2)^2}{16} $&$  
\frac{(1-z^2)^2}{16} $&$  - \frac{1+z^2}{2}$ \\  
$\rho--\mu--\rho--\mu$&$\frac{(3+z^2)^2}{16}  $&$ \frac{(1-z^2)z^2}{8} 
$&$   \frac{(9-z^2)(1+z^2)}{16}$ \\\botrule
\end{tabular}}
\end{table}

The component $d_I$ represents ``the second variation'', or  
one-loop divergence in the one-loop divergence
\begin{eqnarray}
d_I &=& - \frac12 U^{-1}\delta(\delta U) U^{-1} 
 = - \frac12 \Delta (0)  U^{-1}\left\{ \frac12 \partial^2_{\mu}( \delta U )- \right.\nonumber\\
&&\left.-  \frac{3 + z^2}{4}\left[\partial_\mu (\delta U) U^{-1}\partial_\mu U 
- \partial_\mu U U^{-1}\, \delta U \, U^{-1} \partial_\mu U 
+ \partial_\mu U U^{-1}\partial_\mu (\delta U) \right]\right\} U^{-1},\nonumber\\
\delta U &=& \Delta (0)\left[\frac12 \partial^2_{\mu} U - 
\frac{3 + z^2}{4}\partial_\mu U U^{-1}\partial_\mu U\right].
\end{eqnarray}
Therefore it is renormalized away by the redefinition of the $U$ field and
vanishes when the equations of motion are imposed. The counterterms 
renormalizing $U$ field  yield the same expression but twice more and
of the opposite sign. Thus the result is $ - d_I$ 
in correspondence with the results
of \cite{ADE} (for $z=1$ they coincide). 
This is all that was obtained
in that work. In particular no single-pole, $\Delta(0)$
appeared and therefore no new equations were obtained
at the two loop level. Accordingly, the coefficients of the $O(p^4)$ 
coefficients
were deemed to vanish. This will not be the case here. 

The second part represents the remaining terms of 
order $\Delta^2(0)$ in two loop  diagrams 
after subtraction of $d_I$ and it reads
\begin{eqnarray}
d_{II}&=&\frac{1}{32}\Delta^2(0) (1-z^2)
U^{-1} \left\{2(1-z^2)
 \partial_{\rho} U  U^{-1} \partial_\mu\partial_\rho U  U^{-1}
\partial_{\mu} U\right.\nonumber\\ 
&&\left.- (1+z)^2 \partial_{\rho} U  U^{-1} \partial_\mu U  U^{-1} 
\partial_\rho\partial_{\mu} U
- (1-z)^2 \partial_{\rho}\partial_\mu U  U^{-1} \partial_\rho U  U^{-1}
\partial_{\mu} U\right.\nonumber\\ 
&&\left.
+4z^2 \partial_{\rho} U  U^{-1} \partial_\mu U  U^{-1} 
\partial_\rho U  U^{-1}\partial_{\mu} U
\right\} U^{-1}.
\end{eqnarray}
This term is identical, but of opposite sign, to the contributions 
generated by the one-loop 
counterterm in the vertex with two fermions and one boson line,
after its insertion in a one-loop diagram
(see Appendix E). 

To summarize,
we show in Table 1 the distribution of $\Delta^2 (0)$ divergences between
$d_I$ and $d_{II}$. Short-hand notations are used for the corresponding chiral
field (CF) operators, for instance,
 $\mu--\mu\rho--\rho$ corresponds to 
$U^{-1} \partial_{\mu} U  U^{-1} \partial_\mu\partial_\rho U  
U^{-1}\partial_{\rho} U U^{-1}$.

In $d_{III}$ we include those
single-pole divergences, proportional to  
$\Delta (0)$, which are removed once the one-loop
renormalization of $U$ in the finite
nonlocal part of fermion propagator at one loop (see, Eq.~(\ref{1prop})) 
is taken into account,
that is when we replace $U$ by $U+\delta^{(2)}U$ in the one-loop propagator
\begin{eqnarray}
&&\frac14 (1-z^2)\Delta(0)\Delta (A,B)U^{-1}\left[ \partial_\mu (\delta U) U^{-1}
\partial_\mu U \right.\nonumber\\
&&\left.- \partial_\mu U U^{-1}\, \delta U \, U^{-1} \partial_\mu U 
+ \partial_\mu U U^{-1}\partial_\mu (\delta U)
\right] U^{-1}.
\end{eqnarray}
Likewise  in  $d_{IV}$ we include the divergences that are eliminated when
the additional counterterms in the one-boson vertices (those
proportional to $g_i$) are included
in the finite part of the one-loop fermion propagator 
(the terms proportional to $\Delta (A,B)$  in Eq.~(\ref{addi})).\\
One can check that all terms in the two loop fermion propagator linear 
in $\Delta(0)$ and in  $\Delta (A,B)$ belong either 
to  $d_{III}$ or to $d_{IV}$. We present them in Table 2. 
Thus one-loop renormalization  
removes $d_{III}$ and $d_{IV}$ completely.

\vspace{5mm}
\begin{table}[h]
\tbl{Summary of single divergences $\Delta(0)$ which 
are eliminated after the introduction of the one-loop counterterms.} 
{\begin{tabular}{@{}cccc@{}} \toprule
{\bf CF structure}& $d_{III}$& $d_{IV}\leftrightarrow -2 d_{II}$ &
{\bf  Total} \\ \colrule
$\mu^2\rho--\rho$& $\frac{1-z^2}{8}$ &$ 0$ &$ \frac{1-z^2}{8}$   \\  
$\rho--\mu^2\rho$& $ \frac{1-z^2}{8}$& $0$ & $\frac{1-z^2}{8}$   \\  
$\rho--\mu^2--\rho$&$ -\frac{1-z^2}{8} $&$0 $&$ - \frac{1-z^2}{8}$  
\\  
$\rho--\mu--\mu\rho $&$ -\frac{(1-z^2)(3+z^2)}{16}  
$&$  \frac{(1-z^2)(1+z)^2}{16} $&$  - \frac{(1-z^2)(1-z)}{8}$ \\  
$\mu\rho--\mu--\rho$&$ -\frac{(1-z^2)(3+z^2)}{16}
 $&$ \frac{(1-z^2)(1-z)^2}{16} $&$ -\frac{(1-z^2)(1+z)}{8} $ \\  
$\mu\rho--\mu\rho$&$ 0 $&$ 0 $&$  0$
  \\  
$\rho--\mu--\mu--\rho$&$ \frac{(1-z^2)(3+z^2)}{16}$&$ 0 $&$ 
 \frac{(1-z^2)(3+z^2)}{16}$ \\  
$\mu--\mu\rho--\rho $ &$ -\frac{(1-z^2)(3+z^2)}{8} $&$  
-\frac{(1-z^2)^2}{8} $&$  - \frac{1-z^2}{2}$ \\  
$\rho--\mu--\rho--\mu$&$ \frac{(1-z^2)(3+z^2)}{8} $&$ -\frac{(1-z^2)z^2}{4} 
$&$   \frac{(1-z^2)(3-z^2)}{8}$ \\\botrule
\end{tabular}}
\end{table}

Some single-pole divergences remain however. Indeed, there are some
 divergences linear in
 $\Delta(0)$ which come from the double integral 
in the diagrams of Appendix G, (\ref{2-0,8}) and (\ref{2-0,10}),
\begin{eqnarray}
J(A,B)&=&  \int^A_B d\tau_1 \int^{\tau_1}_B d\tau_2   
\partial_{\tau_1}\Delta (\tau_1-\tau_2) \partial_{\tau_2}\Delta 
(\tau_1- \tau_2)\nonumber\\
&=& - \int^{A-B}_0 d\tau \, (A-B -\tau) 
\left[\dot\Delta(\tau)\right]^2. \label{Jab}
\end{eqnarray}
In Appendix H this integral is calculated using two different 
regularizations. The  divergence is found to be
\begin{eqnarray}
d_V &=&  c_V \Delta(0)\left[ U^{-1}
 \partial_{\rho} U  U^{-1} \partial_\mu U  U^{-1} \partial_\mu U  U^{-1}\partial_{\rho} U  U^{-1}\right.\nonumber\\
&&\left.- U^{-1} 
 \partial_{\rho} U  U^{-1} \partial_\mu U  U^{-1} \partial_\rho U  U^{-1}\partial_{\mu} U  U^{-1}\right]\nonumber\\
&\equiv& - \Delta(0) U^{-1} \bar d_V U^{-1}.
\label{dfive}
\end{eqnarray}
with $c_V = \alpha'(1-z^2)^2/8 = \alpha'/2$ for $z = \pm i$.
This term survives after adding all the counterterms and together
with (\ref{dren}) are the only new genuine divergences that can 
contribute to the beta function (single poles).
It must therefore be added to 
the equation of motion at the next order in the $\alpha^\prime$ expansion 
and modifies the term $\delta^{(4)} U$,  $\delta^{(4)}U \rightarrow 
\delta^{(4)} U + 
\bar d_V$ in a crucial manner; 
namely it opens the way to non zero solutions for the coupling constants 
$g_i$ and therefore for nonzero values for the
Gasser-Leutwyler $O(p^4)$ coefficients.

\section{Local integrability and unitarity }

The equation of motion of $O(p^2)$,  Eq.~(\ref{resym}), 
can be obtained from a local action of the Weinberg type (\ref{weinb}), involving 
a unitary matrix $U(x)$, only for $z=\pm i$.
If the corresponding terms with four derivatives that we have just found 
are to be derived from dimension-four operators in a local effective Lagrangian
then certain relations are for sure to be required from the so far 
arbitrary constants $g_{i,r}$.

Such a Lagrangian has only two terms compatible with 
the chiral symmetry if we use the dimension-two equations of motion 
(\ref{resym}), 
\begin{equation}
{\cal L}^{(4)} = \frac12 f_\pi^2 \mbox{tr}\left( K_1 \partial_\mu U \partial_\rho  U^{-1} \partial_{\mu} U \partial_{\rho} U^{-1} 
+ K_2  \partial_\mu U \partial_{\mu} U^{-1} \partial_{\rho} U  \partial_{\rho} U^{-1} 
+ h.c.\right). \label{dim4}
\end{equation}
The terms 
$$\partial_\mu^2 U \partial_\rho  U^{-1} \partial_{\rho} U U^{-1},\qquad  
\partial_\mu^2 U \partial_\rho^2  U^{-1},\qquad  (\partial_\mu^2)^2 U   U^{-1},\qquad  
\partial_\mu\partial_\rho U \partial_{\mu} \partial_{\rho} U^{-1} $$ 
which are in principle possible are reduced to the set (\ref{dim4}) 
with the help  
of integration by parts in the
action and  of  the dimension-two equations of motion (\ref{resym}).

Variation of the previous Lagrangian gives the following 
addition to the equations of motion
\begin{eqnarray}
\frac{\delta S}{\delta U} &=& -  f_\pi^2 U^{-1} \left\{2 K_1\left[
\partial_\mu\partial_\rho U  U^{-1} \partial_{\mu} U  U^{-1} \partial_{\rho} U
+  \partial_\mu U  U^{-1}\partial_\rho U  U^{-1} \partial_{\mu}
\partial_{\rho} U\right.\right.\nonumber\\  
&&\left.\left.
-   \partial_\mu U U^{-1}\partial_\rho U  U^{-1} \partial_{\mu} U U^{-1}
\partial_{\rho} U
-2 \partial_\mu U  U^{-1} \partial_{\rho} U  U^{-1} \partial_{\rho} U U^{-1} 
\partial_\mu U\right.\right.\nonumber\\  
&&\left.\left.
+ \partial_\mu U U^{-1}\partial_\rho^2 U  U^{-1} \partial_{\mu} U \right]
\right.\nonumber\\
&&\left. + K_2\left[\partial_\mu\partial_\rho U  U^{-1} \partial_{\mu} U  
U^{-1} \partial_{\rho} U
+  \partial_\mu U  U^{-1}\partial_\rho U  U^{-1} \partial_{\mu}
\partial_{\rho} U +\right.\right.\nonumber\\  
&&\left.\left.
2 \partial_{\mu} U U^{-1} \partial_\mu\partial_\rho U  U^{-1} 
\partial_{\rho} U\right.\right.\nonumber\\  
&&\left.\left. + \partial_\mu^2 U U^{-1} \partial_{\rho} U U^{-1} 
\partial_{\rho} U
+ \partial_\mu U U^{-1} \partial_{\mu} U^{-1} \partial_{\rho}^2 U
\right.\right.\nonumber\\  
&&\left.\left.
- \partial_\mu U  U^{-1} \partial_{\rho} U  U^{-1} \partial_{\rho} U U^{-1} 
\partial_\mu U-  2 \partial_\mu U U^{-1}\partial_\rho U  U^{-1} 
\partial_{\mu} U U^{-1}\partial_{\rho} U\right.\right.\nonumber\\  
&&\left.\left.
-3 \partial_\mu U U^{-1}\partial_{\mu} U U^{-1} \partial_{\rho} U U^{-1} 
\partial_{\rho} U\right]
\right\}U^{-1}. \label{dimen4}
\end{eqnarray}

\begin{table}[h]
\tbl{Comparison between the 
coefficients of the different chiral field structures  in the equations
of motion derived from a local Lagrangian and from the condition of vanishing
beta function, namely vanishing of the single pole divergences at the
two loop level}
{\begin{tabular}{@{}cccc@{}} \toprule
{\bf CF structure}& $\chi$-lagr.& $d_{g}$ & $d_{V}$ \\ \colrule 
$\mu--\rho--\mu\rho $&$-2(2K_1 + K_2)$&$\frac{1}{16} (1-z^2)(1+z)(g_{1,r} + z g_{2,r})$& 0 \\  
$\mu\rho--\mu--\rho$&$ -2(2K_1 + K_2)$&$\frac{1}{16} (1-z^2)(1-z)(g_{1,r} - z g_{2,r})$& 0\\  
$\mu--\mu\rho--\rho $&$-4K_2$&$\frac18 (1-z^2)(- g_{1,r} + z^2 g_{2,r})$& 0\\ 
$\mu--\rho--\rho--\mu$&$2[(1-z^2) K_1 + K_2] $& 0 &$-c_V$\\ 
$\mu--\mu--\rho--\rho $ &$-2z^2 K_2$& 0 & 0 \\
$\mu--\rho--\mu--\rho$&$4[K_1 + K_2] $&$- \frac14 (1-z^2)z^2g_{3,r}$& 
$c_V$ \\ \botrule
\end{tabular}}
\end{table}

Let us now apply the $O(p^2)$ equations of motion to remove the 
Laplacian on chiral fields 
$\partial_\mu^2 U$. Then
one obtains the set of coefficients for the various chiral field structures
given in Table 3. These coefficients to be determined are then
 compared with the results obtained 
from the coefficients of the one-loop and two-loop single-pole 
divergences (see (\ref{dren}) 
and (\ref{dfive})). For  $z^2 = -1$ only  one solution is possible, implying
\begin{equation}
K_2 = 0, \quad K_1 = -\frac14 c_V = - \frac{\alpha' }{8};\quad\quad
g_{1,r} = - g_{2,r} =  - g_{3,r} = 4 c_V.
\end{equation}
Thus, comparing Eq.~(\ref{dim4}) with the usual parameterization
of the Gasser and Leutwyler Lagrangian\cite{GL},   
\begin{equation}
L_1 = \frac12 L_2 = -\frac14 L_3 = - \frac12 K_1 f_\pi^2 = \frac{ f_\pi^2\alpha' }{16}.
\end{equation}
For $\alpha' = 0.9$ GeV$^{-2}$ and $f_\pi \simeq 93$ MeV it 
yields $L_2 \simeq 0.9 \cdot 10^{-3}$ 
which is quite a satisfactory result\cite{expt}.

So far we have paid no attention to the unitarity of $U$ at the two-loop 
level. Does the variation implied by $d_g$ and $d_V$
respect the perturbative unitarity of $U$? Or does this lead to
a new constraint eventually incompatible with the numerical
previous numerical values? It turns out (see Appendix I) that if 
one accepts arbitrary real coefficients
in the set of dimension-four vertices included in (\ref{dimen4}) then the 
only solution
compatible with the unitarity is given by the parameterization with 
constants $K_1$ and
$K_2$. Thus the requirement to preserve unitarity under field 
renormalization is entirely
equivalent to the local integrability condition, similarly to the case
of dimension-two operators. This is a remarkable result that hints to the
consistency of the whole procedure.

\section{Conclusions: conformal invariance and all that}

Conformal invariance is a subtle issue. As we have discussed in
detail in the introduction this is a necessary ingredient for 
the consistency of string theory and it is, according to 
general principles of field theory, equivalent to requiring the vanishing
of the beta functions.

There are several sources of breaking of conformal invariance 
in the present model. We have been concerned with the equations
of motion for the $U(x)$ string field functional. The beta functional
has been computed up to two loops, properly including the 
unitarity constraints. The vanishing of the single pole divergences
at one loop 
leads to the familiar non-linear sigma model. At two
loops the $O(p^4)$ terms of Gasser and Leutwyler appear. The
numerical
values turns out to be identically determined by the arguments
of locality and unitarity and they therefore determine the
`renormalized' value of $g_{i,r}$ in terms of the constant $c_V$.
The latter being the coefficient of a single loop divergence
in an integral appears to be manifestly universal. The former
could be in principle ambiguous since a change in the renormalization
scheme (for instance by choosing the subtraction to a value different
from
$\mu \tau= 1$ in the $x$-propagator) may shift a finite 
piece between $\Delta(0)$ and the
renormalized
value of the coupling.

However, this ambiguity is of no relevance. The requirements
of locality and unitarity seem to restrict the `measurable' value 
of $g_{i,r}$
to specific, well determined values. If one changes the
renormalization scheme by a finite amount $\delta$, the `measurable' 
couplings will all change by the same amount to $\bar{g}_{i,r}=
g_{i,r}+\delta$, but it is this last quantity the one that will then
appear in the equations of motion and will be related to $c_V$, which
is universal, so this
change of scheme is immaterial.   

Furthermore, the renormalized couplings $g_{i,r}$ are conformal
factor independent, so the relations determined in the
previous section hold in any conformal frame and so are the
values for the $O(p^4)$ coefficients of the chiral Lagrangian.
This is of course in full agreement with
general considerations based in QCD and chiral Lagrangians (in the
large $N$ limit, that is, where the string picture is supposed
to hold).

In \cite{ADE} a general, but tentative, 
argument (based in conformal invariance) was
given that, if valid, would imply the vanishing of the  $O(p^4)$
coefficients to all orders in the inverse string tension.
Here the argument simply fails because the additional couplings
which are required to renormalize the model are dimensional.
These couplings were not considered in \cite{ADE} because the need
for additional counterterms was not manifest along the renormalization
process due to the (incorrect) choice of the value of $z$.

Of course this immediately raises a new issue. The new counterterms
seem to entail a breakdown of `classical' conformal
invariance. Indeed, introducing the coupling $g_i$, which are of dimension
$M^{-2}$ makes the action non-conformally invariant already at the
classical level, in a way that is similar
to the introduction of a coupling to the tachyon in the familiar bosonic
string theory. Although the issue is somewhat independent of the
equations of motion for the matrix field $U(x)$ we have to give
an  answer to this difficulty if conformal invariance is to hold.

To demand
 conformal invariance in the string-mode sector, we have
to guarantee the vanishing of the trace of the energy momentum tensor.
This issue can be discussed in several ways, but let us proceed to
determine the trace of the energy momentum tensor. On general grounds
we can write
\begin{equation}
\Theta= \sum_i \beta_i O_i,
\end{equation}
where $\beta_i$ is the beta function for the $i$-th coupling and 
$O_i$ the accompanying operator. Then, when computing the trace of the
energy momentum tensor, 
the part of the tensor  bilinear in fermion fields 
leads to a contribution that contains operators of the form
(\ref{count}) with their couplings $g_i$ replaced by the appropriate beta
 functions. These beta functions contain a classical, purely
 engineering, part as well as a quantum one, which has been computed
in the text, and which is the same for all couplings $g_i$.
In the string-mode sector their contribution is given by
the averaging over the fermion vacuum and it vanishes 
according to the arguments
presented in the Appendix D. There is no a fermion induced string action
when the $CP$ symmetry (\ref{CP1}) holds. Thus the presence of conformally 
non-invariant interaction on the boundary does not affect conformal
symmetry in the bulk. 

Throughout this paper we have systematically used conventional
perturbation theory and the fermion loop expansion which however may
be more cumbersome when one proceeds to the vertices with larger number
of emitted bosons and to higher orders in the loop expansion. As
an alternative, direct functional methods can be developed, an example
of which is displayed in the Appendix J. This approach can be also exploited
to show how to go ahead without fermions and so bypassing the
potential problems related to breakdown of conformal
anomaly since it allows a formulation where the `quarks' are
integrated
out. Doing so apparently removes any conformal anomaly present in the 
boundary as we just discussed.

On the other hand,
at some point we do expect a genuine breakdown of conformal invariance. This
is so because of the conformal anomaly that the bosonic string theory
must necessarily exhibit. In fact, this breakdown (tantamount to
introducing a scale dependence) is quite welcome as we do not
expect the simple string picture presented here to be 
even approximately true at very
short distances (even within the large $N$ limit). The naive bosonic string
action used in the present paper does not prevent large Euclidean 
world sheets from crumpling \cite{pol1}. It does not also describe correctly
the high-temperature behavior of large $N$ QCD \cite{polch}.

According to by now common lore\cite{alv}, the dependence 
on the conformal factor has to be understood in terms of the
renormalization group flow.
This dependence has not been worked out in detail, but it would
be feasible to take it into account perturbatively in a way
not too different from standard sigma model calculations in string
theory by including dilaton, spin two, etc. backgrounds. The
dilaton beta function precisely contains the $d-26$ term 
characteristic of the anomaly. At some order in the
$\alpha^\prime$ and topological expansion there would appear a coupling 
between the dilaton and the chiral fields through the exchange
of $x$ string variables. It would be interesting how the 
hadronic degrees of freedom can modify the anomaly condition, if at all.

A QCD induced string action may also include 
nonlocal \cite{pol2,quev,ant}  or, at least, higher-derivative
\cite{diam} vertices breaking manifestly conformal symmetry
which help to make the strings smooth and supply the correct
high-temperature asymptotics. However, as we are concerned here
with the low-energy string properties we do not expect that the strategy
and technique to derive the chiral field action needs any significant
changes to be adjusted to a modified QCD string action. The only changes
may come out from the short distance behavior of the 
modified string propagator.

Finally we enumerate some of the simplifications and missing points of
the approach undertaken in the present paper.

\noindent
1) From the beginning we have restricted ourselves to the $SU(2)$ 
global flavor
group. The reason is that only parity-even terms in the equations 
of motion can be
revealed from the simple fermion Lagrangian (\ref{lag}). As it is supposed that
all relevant meson degrees of freedom are reproduced by the hadronic string
one cannot expect the appearance of multi-fermion
interaction which may effectively arise only due to heavy-mass reduction
of glueballs and hybrids suppressed in the large $N$ limit.
Then the only way to obtain the parity-odd Wess-Zumino-Witten terms is
to supplement one-dimensional fermions with spinor degrees of freedom,
 i.e., for instance, to add reparameterization invariant vertex
\begin{equation}
\Delta L_f =\xi \bar\psi \gamma^\mu \dot{x}_\mu\psi, 
\end{equation}
with Dirac matrices $ \gamma^\mu$ and new dimensional constant $\xi$.
This extension will be investigated elsewhere.

\noindent
2) One can easily add external electromagnetic fields and thereby
calculate also the chiral constants $L_9, L_{10}$. However it is not yet
clear how to introduce current quark masses consistently with the
open string picture with ultra-relativistic quarks at their ends.
Hence the other chiral constants need more efforts to be understood
within the string approach if the boundary fermions were indeed
associated with quarks.

\noindent
3) Going back to the possibility of higher derivative \cite{diam} or nonlocal
\cite{pol2,quev,ant}
or five-dimensional \cite{pol3,alv} action
for the hadronic string, we find it interesting to explore the traces
of deviations from the conformal string theory in low-energy
chiral constants and phenomenology. 
Including the running of the $O(p^4)$ coefficients
would however require the introduction of $1/N$ corrections. We do not know
if the string picture can be consistently implemented when subleading
corrections are taken into account.

\noindent
4) As we have mentioned in this section ---and emphasize here once
more--- the conformal anomaly can hopefully be taken into account consistently
within the present approach, but he have not really attempted to 
do so in this present work. This is a major challenge that
we leave for the future. 

\noindent
5) The value for the coefficients of $O(p^4)$ that we have found
are, up to the usual factors of $\pi$, etc.  rational numbers.
In other words, they are not the first terms in the Taylor 
expansion of a $\Gamma$ function or similar transcendental 
functions. What would be the appropriate amplitude replacing
Veneziano formula then when the proper symmetries of the QCD vacuum 
are considered such as in the present case? We do not know.

All things considered, we believe that the objectives we set in
the introduction have been achieved. We have obtained the 
leading effective action for low energy QCD from two very simple 
requirements: chiral invariance and conformal symmetry and
two simple ideas: perturbing about the true vacuum of QCD and
using the simplest possible effective world-sheet action. 
The approach seems to be oversimplified, but has proven to be considerably
robust. All cross checks and physical requirements are met.
The outcome is very sensible from a phenomenological point
of view. Perhaps more importantly, several interesting
avenues are open for future exploration and a fully consistent
approach may be developed. 
We also hope that the considerations presented in this
work may be of use to the ample string community since some
of the techniques we have employed appear to be relevant in
a wider context.

\bigskip\bigskip

\section*{Acknowledgments}
\bigskip

We thank
A. Dobado who participated in the initial stages of this work for multiple
discussions and G.Venturi for invaluable comments. This work
was finished during the visit of one of the authors (D.E.)
to the Departamento de F\'\i sica of the
Universidad Cat\'olica de Chile, whose hospitality
is gratefully acknowledged.
We acknowledge the financial support from
MCyT FPA2001-3598 and CIRIT 2001SGR-00065 and the European
Networks EURODAPHNE and EUROGRID and also from the project
Fondecyt 7010967. The work of J.A.
has been partially supported by Fondecyt 1010967 and L.B. has been supported
by Fondecyt 2000028. J.A. and L.B. wish to thank the warm hospitality
extended to them
during their visit to the Universitat de Barcelona.
A.A. is partially supported by Grant RFBR
01-02-17152, Russian
Ministry of Education Grant E00-33-208 and by The Program {\sl Universities
of Russia: Fundamental Investigations} (Grant 992612). By this work A.A. and 
D.E. contribute 
to the fulfillment of  Project INTAS 2000-587.
\vfill
\eject

\vfill\eject

\appendix
\section{}

To construct the most general reparameterization- and chirally invariant 
action on the boundary of the string one can use the following
set of operators bilinear in fermion variables and of the minimal
dimension one
\begin{eqnarray}
&&\bar\psi_L U  \dot\psi_R,\quad \dot{\bar\psi_L} U \psi_R, \quad \dot{\bar\psi_R} U^+ \psi_L,\quad
\bar\psi_R U^+ \dot\psi_L,\nonumber\\
&& \bar\psi_L \dot\psi_L,\quad \bar\psi_R   \dot\psi_R ,\quad \bar\psi_L\dot{U} U^+ \psi_L,\quad
\bar\psi_R U^+ \dot{U} \psi_R. \label{vertex}
\end{eqnarray}
Other vertices like $\bar\psi_L\dot{U} \psi_R$ can be decomposed in a linear combination 
of basic vertices (\ref{vertex}) after integration 
by parts in the action  $S_b = \int d\tau L_f$. 
The multi-fermion local interaction is suppressed in the leading large $N$
approximation as we assume that
all meson degrees of freedom relevant in this limit are reproduced by the
hadronic string. Then a multi-fermion
interaction may effectively arise only due to heavy-mass reduction
of glueballs, hybrids and multiquark mesons suppressed in the large $N$
limit.

As the case of $CP$ symmetric action is of the most importance to provide the conformal symmetry
 we restrict ourselves with the analysis of this action only.
Thus the general $CP$ invariant hermitean Lagrangian takes the form:
\begin{eqnarray}
L_f&=&b \bar\psi_L U  \dot\psi_R + b^* \dot{\bar\psi_L} U \psi_R + b^* \dot{\bar\psi_R} U^+ \psi_L +
b \bar\psi_R U^+ \dot\psi_L \nonumber\\
&&+ i c\left( \bar\psi_L \dot\psi_L + \bar\psi_R  \dot\psi_R\right) + 
i d \left(\bar\psi_L\dot{U} U^+ \psi_L - \bar\psi_R U^+ \dot{U} \psi_R\right),\label{lagra}
\end{eqnarray}
where the constant $b$ is complex whereas the constants $c,d$ are real.
As compared to the minimal Lagrangian (\ref{lag}) the general $L_f$ contains three more real parameters,
$\mbox{\rm Im}(b), c,d$. Now let us show that by the local rotation of fermion variables preserving their chiral structure,
\begin{eqnarray}
 \psi_L &=& \alpha_1 \psi_L + \alpha_2 U \psi_R, \quad\bar\psi_L = \alpha_1^* \bar\psi_L + \alpha_2^* \bar\psi_R U^+,\nonumber\\
\psi_R &=& \beta_1 \psi_R + \beta_2 U^+ \psi_L,\quad \bar\psi_R = \beta_1^* \bar\psi_R + \beta_2^* \bar\psi_L U, \label{rot}
\end{eqnarray}
one can eliminate the redundant vertices and reduce the Lagrangian (\ref{lag}) to the minimal form.
In order to prove it we transform the minimal Lagrangian (\ref{lag})  with the help of rotation (\ref{rot}).
The initial set of constants is $b = (a + i)/2, c = d = 0$. The final set of vertices will be $CP$ invariant
if the following conditions are fulfilled:
\begin{eqnarray}
 \alpha_1\beta_1^* &=& \alpha_1^*\beta_1,\\  \alpha_2\beta_2^* &=& \alpha_2^*\beta_2,\\
 \alpha_1 \beta_2^* &=& \alpha_2^* \beta_1. \label{condi}
\end{eqnarray} 
The first two constraints relate some phases,
\begin{eqnarray}
 \alpha_1 &=& |\alpha_1| \exp(i\phi_1),\quad \beta_1 =  \pm |\beta_1|\exp(i\phi_1),\nonumber\\
 \alpha_2 &=& |\alpha_2| \exp(i\phi_2),\quad \beta_2 =  \pm |\beta_2|\exp(i\phi_2),
\end{eqnarray} 
whereas the third one eliminates one of the moduli of $ |\alpha_j|$ and $|\beta_j|$. Three remaining
moduli and the relative phase $\phi_1 - \phi_2$ turn out to be sufficient to fit three real constants $\mbox{\rm Im}(b), c,d$,
\begin{eqnarray}
|\alpha_1||\beta_2| &=& |\alpha_2||\beta_1| = \frac12 \sqrt{c^2 + \frac{(2d + c)^2}{a^2}} \equiv \frac12 \zeta,\nonumber\\
\cos(\phi_1 - \phi_2) &=& \frac{c}{\zeta},\nonumber\\
\pm |\alpha_1||\beta_1| \pm |\alpha_2||\beta_2| &=&\frac12 \zeta \left( 
\pm \frac{|\beta_1|}{\beta_2|}\pm \frac{|\beta_2|}{|\beta_1|}\right) = \frac12 \mbox{\rm Im}(b). 
\end{eqnarray}
Evidently this system of equations has solutions for arbitrary $\mbox{\rm Im}(b), c,d$ and therefore 
the minimal Lagrangian can be always obtained by an equivalence transformation (\ref{rot}) of fermion fields.
At the quantum level this local transformation does not yield a nontrivial Jacobian  when one
applies the dimensional regularization to calculate it.

\begin{figure}
\centerline{\psfig{file=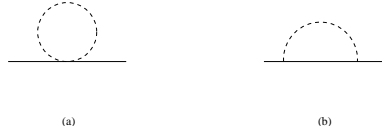,width=5cm}}
\vspace*{8pt}
\caption{One-loop diagrams for the propagator.}
\end{figure}

\section{}
In this and the following appendixes we present the results of our
perturbative calculation. When necessary, both finite and divergent
parts are given.
Diagrams are labeled according to the figure number.

Diagram 1.a:
\begin{equation}
- \frac12 \theta(A - B) \Delta (0) U^{-1} 
\partial^2_{\mu} U  U^{-1}.
\end{equation}

Diagram 1.b:
\begin{equation}
\frac14 \theta(A - B) \left[(3 + z^2)\Delta (0) + (1-z^2)\Delta (A,B)\right] U^{-1} 
\partial_\mu U U^{-1}\partial_\mu U U^{-1}.\label{1prop}
\end{equation}

\section{}
In this appendix we show the result of the one-loop calculation of the
vertex with two fermions and one boson line. The divergences that
appear are not fully eliminated by a redefinition of $U$ and additional
counterterms with higher derivatives are called for.
\begin{figure}
\centerline{\psfig{file=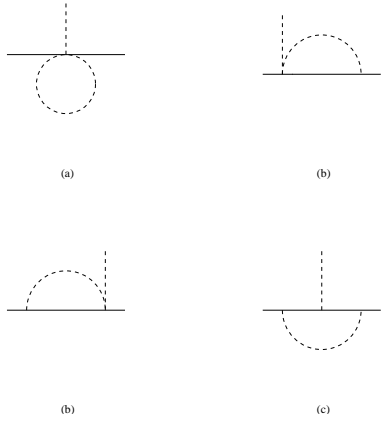,width=5cm}}
\vspace*{8pt}
\caption{One-loop diagrams for the vertex with one $\tilde x$-field.}
\end{figure}

Diagram 2.a:
\begin{equation}
- \frac14 \theta(A - B) \Delta (0) U^{-1} 
\partial_{\mu}(\partial^2 U)  U^{-1} \left[\bar x_\mu(A) +\bar x_\mu(B) +
z\left(\bar x_\mu(A) -\bar x_\mu(B)\right)\right] .
\end{equation}

Diagram 2.b:
\begin{eqnarray}
&&\frac14 \theta(A - B) U^{-1} 
\left\{\left[\Delta (0)
\left(2 (1+ z)\bar x_\mu(A) + (1 - z)^2\bar x_\mu(B)\right)\right.\right.\nonumber\\ 
&&\left.\left.
+ \Delta (A,B) (1-z^2)\bar x_\mu(B) \right]
\partial_\nu U U^{-1}\partial_\mu\partial_\nu U\right.\nonumber\\
&&\left. +
 \left[\Delta (0)\left((1+ z)^2\bar x_\mu(A) + 
2(1 - z)\bar x_\mu(B)\right)\right.\right.\nonumber\\ 
&&\left.\left. + \Delta (A,B) (1-z^2)\bar x_\mu(A) \right]
\partial_\mu \partial_\nu U U^{-1}\partial_\nu U \right\}U^{-1}.
\end{eqnarray}

Diagram 2.c:
\begin{eqnarray}
&&- \frac18 \theta(A - B) \left\{\Delta (0) 
\left[(1+ z)^2 (3 - z) \bar x_\mu (A) + (1 - z)^2 (3 + z)\bar x_\mu (B)\right]\right.\nonumber\\
&&\left. +  \Delta (A,B) (1-z^2)\left[\bar x_\mu(A)(1+z) + \bar x_\mu(B)(1-z)\right]\right.\nonumber\\
&&\left. +(1-z^2) \int^A_B d\tau \dot{\bar x}_\mu(\tau) 
\left[\Delta(\tau, B)(1-z) - \Delta(A,\tau)(1+z)\right]\right\}\nonumber\\
&& \times U^{-1} \partial_\nu U U^{-1} \partial_\mu U U^{-1} \partial_\nu U U^{-1}.
\end{eqnarray}

\section{}
This appendix has to do with the fermion determinant and the 
vanishing scale anomaly discussed in the main body of the paper.

There are two mutually conjugated operators in the bilinear form of 
the Lagrangian (\ref{lag}):
\begin{eqnarray}
{\cal D} &=& \frac{i}{2}\left[ (1 - z) U(\tau) i \partial_\tau + (1 + z) i \partial_\tau\biggl( U(\tau) \right],\nonumber\\
{\cal D}^\dagger &=& \frac{i}{2}\left[ (1 + z^*) U^+(\tau) i \partial_\tau + (1 - z^*) i \partial_\tau\biggl( U^+(\tau) \right]
\label{oper}.
\end{eqnarray}
Therefore the total fermion determinant
(the result of integration over fermions) or fermion loop contribution 
can be represented by
\begin{equation}
Z_f = \left|\!\left|{\cal D}{\cal D}^\dagger \right|\!\right|
= \left|\!\left| (i \partial_\tau - \frac{i}{2}(1 - z) \dot{U} U^+ )  (i \partial_\tau - \frac{i}{2}(1 - z^*) \dot{U} U^+ ) \right|\!\right|,
\end{equation}
where we have restricted ourselves with unitary fields $U$.
Now one can factorize out the infinite constant for free operators and find the nontrivial part in terms of
fermion propagators:
\begin{eqnarray}
Z_f &=& \left|\!\left| (i \partial_\tau)^2 \right|\!\right|  
\exp\left\{\mbox{\rm Tr}
\left(\log\left(1- \frac{i}{2i \partial_\tau}(1 - z) \dot{U} U^+\right)
\right.\right.\nonumber\\ 
&&\left.\left.+ 
\log\left(1- \frac{i}{2i \partial_\tau}(1 - z^*) \dot{U} U^+\right)\right)
\right\}\nonumber\\
&=& C \exp\left\{- \theta(0)\frac12 \int^\infty_{-\infty} d\tau (1 - z+ 1 - z^*)\mbox{\rm tr}(\dot{U} U^+) \right\}\nonumber\\
&=& C \exp\left\{- \frac12 \int^\infty_{-\infty} d\tau 
\mbox{\rm tr}(\dot{U} U^+) \right\}. \label{deter} 
\end{eqnarray}
Herein the triangle property of the free fermion propagator has been exploited,
\begin{equation}
\langle\tau|\frac{i}{i \partial_\tau + i\epsilon} |\tau'\rangle
 =\theta(\tau - \tau'),
\end{equation}
which follows from the advanced Green function prescription by adding $+ i\epsilon$. As a result,
only the first order in the expansion of the logarithms in (\ref{deter}) survives when the functional
trace operation is performed. The value $ \theta(0) = 1/2$ and the
$CP$
 invariant choice
$z = - z^*$ are employed. More rigorously this result can be obtained by taking the finite proper-time interval
with anti-periodic boundary conditions for fermion fields. Then the determinant will be given by the product of
discrete eigenvalues of the operators (\ref{oper}). When proceeding to the infinite line limit and for the advanced
prescription with $+ i\epsilon$ one recovers exactly the functional presented in the last line of (\ref{deter}).

Evidently, for $SU(N)$ groups and for $U(1)$ groups with periodic boundary conditions the exponent in
(\ref{deter}) vanishes and therefore the fermion loop contribution from the minimal Lagrangian is absent. It automatically eliminates
the scale anomaly $\sim \langle L_f\rangle_{vac}$ and thereby the conformal symmetry remains intact by vacuum polarization
effects.

For the extended Lagrangian including the higher dimensional vertices 
(\ref{count}) the derivation of the
fermion determinant is similar. The corresponding differential operators look as follows:
\begin{eqnarray}
\tilde{\cal D} &=& \left(i \partial_\tau - \frac{i}{2}(1 - z) \dot{U} U^+ 
+ \frac{i}{8} (1 - z^2)\left[ (g_1 - z g_2)
\partial_\nu \dot{U} U^+\partial_\nu U U^+ (g_1 + z g_2)  \right.\right.\nonumber\\
&&\left.\left. - \partial_\nu U U^+\partial_\nu \dot{U}U^+ 
+ 2zg_3 \partial_\nu U U^+ \dot{U} U^+ \partial_\nu U U^+
\right]\right) U,\nonumber\\
\tilde{\cal D}^\dagger &=& U^+ \left(i \partial_\tau - \frac{i}{2}(1 - z^*) \dot{U} U^+ - \frac{i}{8} (1 - (z^*)^2)\left[  (g_1 - z^* g_2)
U \partial_\nu U^+ U \partial_\nu \dot{U}^+\right.\right.\nonumber\\
&&\left.\left.
 - (g_1 + z^* g_2)
U\partial_\nu \dot{U}^+ U\partial_\nu U^+  + 
2z^* g_3 U \partial_\nu U^+ U\dot{U}^+  U \partial_\nu U^+
\right]\right).
\label{oper2}
\end{eqnarray}
Again only the first order in the expansion of the logarithmic trace is not vanishing:
\begin{eqnarray}
\log Z_f &=& C' + \frac{1}{8} \int^\infty_{-\infty} d\tau \mbox{\rm tr}
\left((1 - z^2)\left[ 
z g_2 \partial_\nu \dot{U} \partial_\nu U^+  + 
z g_3 \partial_\nu U U^+ \dot{U} U^+ \partial_\nu U U^+
\right]\right.\nonumber\\
&&\left. + (1 - (z^*)^2)\left[-  z^* g_2
\partial_\nu \dot{U} \partial_\nu U^+ + 
z^*  g_3 \partial_\nu U U^+ \dot{U} U^+ \partial_\nu U U^+
\right]\right) .
\end{eqnarray}
The vertices proportional $g_1$ do not appear after tracing. Let us
take the 
$CP$ symmetric constants $z = - z^*$.
The vertices with the coupling constant $g_3$ happen to be proportional to $z + z^*$ and therefore
vanish.  As to the $g_2$ terms they form a total time derivative in
the $CP$ invariant case,
\begin{equation}
\log Z_f = C' +  \frac{1}{8} \int^\infty_{-\infty} 
d\tau (1 - z^2) z g_2 \ \partial_\tau \mbox{\rm tr}( 
\partial_\nu U \partial_\nu U^+) = 0,
\end{equation}
the latter taking place for periodic boundary conditions.

Thus, in spite of the fact that the higher dimensional vertices bring a scale
dependence into the Lagrangian (the constants $g_j$ are dimensional), 
they do not generate
a scale anomaly due to fermion loops iff the $CP$ symmetry is imposed on the
Lagrangian.

\section{}

Here we list the contribution from the additional counterterms
(\ref{adddiv}) to the fermion propagator.
\begin{figure}
\centerline{\psfig{file=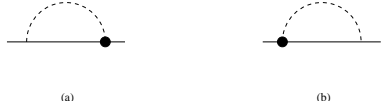,width=5cm}}
\vspace*{8pt}
\caption{One-loop counterterms to the fermion propagator coming
 from the additional vertices.}
\end{figure}

Diagrams 3: 
       
\begin{eqnarray}
&&\frac14 \left( \Delta (0) - \Delta (A,B)\right) \left[(1+z)U^{-1} \partial_\mu U  U^{-1} \phi_\mu U^{-1}\right.\nonumber\\
&&\left.
- (1-z) U^{-1} \phi_\mu U^{-1} \partial_\mu U  U^{-1}\right], \label{addi}
\end{eqnarray}
where
\begin{eqnarray}
\phi_\mu &=& \Delta (0) \frac{1 - z^2}{2}\left(  \frac{1 - z}{2}
\partial_\mu \partial_\nu U U^{-1}\partial_\nu U 
 -  \frac{1 + z}{2}
\partial_\nu U U^{-1}\partial_\mu\partial_\nu U \right.\nonumber\\
&&\left.+ z \partial_\nu U U^{-1} \partial_\mu 
U U^{-1} \partial_\nu U\right) . 
\end{eqnarray}
When retaining only the terms $\sim \Delta^2 (0)$  
one reproduces exactly $2 d_{II}$.

\section{}
Here we list the diagrams that are relevant to the calculation of the
one-loop vertex involving two fermions and two boson lines.

Diagram 4.a:
\begin{equation}
- \frac18 \theta(A - B) \Delta (0)\left[\bar x_\mu(A) \bar x_\nu(A) (1 + z) 
+ \bar x_\mu(B) \bar x_\nu(B) (1 - z)\right] U^{-1} 
  \partial_\mu\partial_\nu \partial^2_{\rho} U  U^{-1}.
\end{equation}

Diagram 4.b:

Its divergent and finite parts are
\begin{eqnarray}
&& \frac18 \theta(A - B) \left\{ \Delta (0)\left[\bar x_\mu(A) \bar x_\nu(A)2 (1 + z) 
+ \bar x_\mu(B) \bar x_\nu(B) (1 - z)^2\right]\right.\nonumber\\
&&\left. + \Delta(A,B)\bar x_\mu(B) \bar x_\nu(B) (1-z^2)\right\} U^{-1} 
 \partial_{\rho} U  U^{-1}  \partial_\mu\partial_\nu \partial_{\rho} U  U^{-1}.
\end{eqnarray}

\begin{figure}
\centerline{\psfig{file=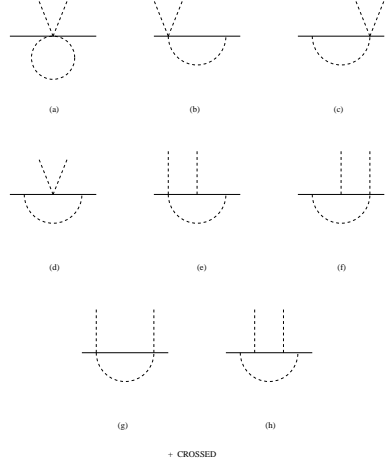,width=5cm}}
\vspace*{8pt}
\caption{One-loop diagrams for the vertex with two $x$-fields.}
\end{figure}

Diagram 4.c:

Its divergent and finite parts are
\begin{eqnarray}
&& \frac18 \theta(A - B)\left\{\Delta (0)\left[\bar x_\mu(A) \bar x_\nu(A) (1 + z)^2 
+ \bar x_\mu(B) \bar x_\nu(B) 2 (1 - z)\right]\right.\nonumber\\
&&\left. + \Delta(A,B)\bar x_\mu(A) \bar x_\nu(A) (1-z^2)\right\} U^{-1} 
  \partial_\mu\partial_\nu \partial_{\rho} U  U^{-1} \partial_{\rho} U  U^{-1}.
\end{eqnarray}

Diagram 4.d:

Its divergent and finite parts are
\begin{eqnarray}
&&- \frac{1}{16} \theta(A - B)\left\{ \Delta (0)\left[\bar x_\mu(A) \bar x_\nu(A) (1 + z)^2(3 - z) 
+ \bar x_\mu(B) \bar x_\nu(B) (1 - z)^2 (3 + z)\right]\right.\nonumber\\
&&\left. +  \Delta(A,B)(1-z^2) \left[\bar x_\mu(A) 
\bar x_\nu(A) (1+z) + \bar x_\mu(B) \bar x_\nu(B) (1-z)\right]\right.\nonumber\\
&&\left. +  (1-z^2) \int^A_B d\tau (\bar x_\mu \dot{\bar x}_\nu (\tau)+ \dot{\bar x}_\mu \bar x_\nu (\tau))
\left[ \Delta (\tau, B)(1 - z) - \Delta (A, \tau)(1 + z)\right]\right\}\nonumber\\
&&\times U^{-1} 
 \partial_{\rho} U  U^{-1}  \partial_\mu\partial_\nu U  U^{-1} \partial_{\rho} U  U^{-1}.
\end{eqnarray}

Diagram 4.e:

Its divergent and finite parts are
\begin{eqnarray}
&&-  \frac{1}{8} \theta(A - B)\left\{ 
\Delta (0)\left[\bar x_\mu(A) \bar x_\nu(A) (1 + z)^2 (3 - z) 
+ \bar x_\mu(B) \bar x_\nu(B) (1 - z)^2 (3+z)\right.\right.\nonumber\\
&&\left.\left. + (1 - z)^2 (1+z)
\int^A_B d\tau \bar x_\mu(\tau) \dot{\bar x}_\nu(\tau)\right]\right.\nonumber\\
&&\left. + \Delta (A,B) (1-z^2)\left[\bar x_\mu(A) \bar x_\nu(B) (1 + z) 
+ \bar x_\mu(B) \bar x_\nu(B) (1 - z)\right]\right.\nonumber\\
&&\left. + (1-z^2)\int^A_B d\tau \left[ \dot{\bar x}_\mu(\tau ) \bar x_\nu (B)
\Delta (\tau, B)(1 - z) -  \dot{\bar x}_\mu(\tau ) 
\bar x_\nu (\tau)\Delta (A, \tau)(1 + z)\right]\right\}\nonumber\\
&&\times  U^{-1} \partial_{\rho} U  U^{-1}
 \partial_{\mu} U  U^{-1}  \partial_\nu \partial_{\rho} U  U^{-1}.
\end{eqnarray}

Diagram 4.f: 

Its divergent and finite parts are
\begin{eqnarray}
&&- \frac{1}{8} \theta(A - B) \left\{ \Delta (0)\left[\bar x_\mu(A) \bar x_\nu(A) (1 + z)^2(3 - z) 
+ \bar x_\mu(B) \bar x_\nu(B) (1 - z)^2 (3+z)\right.\right.\nonumber\\
&&\left.\left. -  (1 - z) (1+z)^2
\int^A_B d\tau \dot{\bar x}_\mu(\tau) {\bar x}_\nu(\tau)\right]\right.\nonumber\\
&&\left. + \Delta (A,B) (1-z^2)\left[\bar x_\mu(A) \bar x_\nu(A) (1 + z) 
+ \bar x_\mu(A) \bar x_\nu(B) (1 - z)\right]\right.\nonumber\\
&&\left. + (1-z^2)\int^A_B d\tau \left[ \bar x_\mu(\tau ) \dot{\bar x}_\nu (\tau)
\Delta (\tau, B)(1 - z) -  \bar x_\mu(A)  \dot{\bar x}_\nu (\tau)
\Delta (A, \tau)(1 + z)\right]\right\}\nonumber\\
&&\times  U^{-1} 
 \partial_{\rho} \partial_\mu U  U^{-1} \partial_\nu U  U^{-1}\partial_{\rho} U  U^{-1}.
\end{eqnarray}

Diagram 4.g:

Its divergent and finite parts are
\begin{eqnarray}
&& \frac{1}{8} \theta(A - B)\left\{ \Delta (0)\left[\bar x_\mu(A) \bar x_\nu(A) (1 + z)(3+z)) 
+ \bar x_\mu(B) \bar x_\nu(B) (1 - z) (3-z)\right.\right.\nonumber\\
&&\left.\left. + (1-z^2) \int^A_B d\tau \left(\bar x_\mu (\tau)\dot{\bar x}_\nu (\tau) 
- \dot{\bar x}_\mu (\tau)\bar x_\nu(\tau)\right) \right]\right.\nonumber\\
&&\left. + 2 \Delta (A,B) (1-z^2)\bar x_\mu(A) \bar x_\nu(B)\right\}\nonumber\\  
&&\times U^{-1} 
 \partial_{\rho} \partial_\mu U  U^{-1} \partial_{\rho} \partial_\nu U  U^{-1}.
\end{eqnarray}

Diagram 4.h:

Its divergent and finite parts are
\begin{eqnarray}
&& \frac{1}{16} \theta(A - B) \left\{\Delta (0)\left[\bar x_\mu(A) \bar x_\nu(A)2 (1 + z)^2(3 - z) 
+ \bar x_\mu(B) \bar x_\nu(B)2 (1 - z)^2 (3+z)\right]\right.\nonumber\\
&&\left. + \Delta (A,B)(1-z^2)\left[\bar x_\mu(A) \bar x_\nu(A)2 (1 + z) 
+ \bar x_\mu(B) \bar x_\nu(B) (1 - z)^2\right.\right.\nonumber\\ 
&&\left.\left. +
 \bar x_\mu(A) \bar x_\nu(B)(1 - z^2) - (1-z^2)\int^A_B d\tau  \dot{\bar x}_\mu(\tau) \bar x_\nu(\tau)\right]\right.\nonumber\\
&&\left. + (1-z^2)\int^A_B d\tau \left[\Delta (\tau,B) \left( \bar x_\mu(\tau) \dot{\bar x}_\nu(\tau) 2 (1-z)\right.\right.\right.\nonumber\\
&&\left.\left.\left. + 
 \dot{\bar x}_\mu (\tau)\bar x_\nu(\tau) 
(1-z^2)
+ \dot{\bar x}_\mu (\tau)\bar x_\nu(B)(1-z)^2 \right)\right.\right.\nonumber\\ 
&&\left.\left. - 
\Delta (A,\tau) \left( \bar x_\mu(\tau) \dot{\bar x}_\nu(\tau) (1-z^2) + 
 \dot{\bar x}_\mu (\tau)\bar x_\nu(\tau) 2(1+z)+ \bar x_\mu(A) \dot{\bar x}_\nu(\tau) (1+z)^2 \right)\right]\right.\nonumber\\ 
&&\left. - (1-z^2)^2\int^A_B d\tau_1 \int^{\tau_1}_B d\tau_2   
\dot{\bar x}_\mu (\tau_1) \dot{\bar x}_\nu(\tau_2)\Delta (\tau_1,\tau_2)\right\}\nonumber\\
&&\times U^{-1} 
 \partial_{\rho} U  U^{-1} \partial_\mu U  U^{-1} \partial_\nu U  U^{-1}\partial_{\rho} U  U^{-1}.
\end{eqnarray}

All divergences are removed by combining the renormalization of $U$
and the additional counterterms determined from the one loop vertex
with one external boson lines. No new counterterms are required.

\section{}
Two loop diagrams for the fermion propagator. Only the
divergent parts are needed.

\begin{figure}
\centerline{\psfig{file=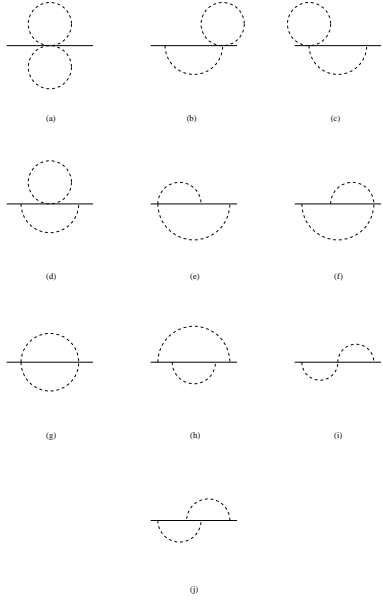,width=5cm}}
\vspace*{8pt}
\caption{Two-loop diagrams for the propagator.}
\end{figure}

Diagram 5.a:

Its contribution is fully divergent,
\begin{equation}
- \frac18 \theta(A - B) \Delta^2 (0) U^{-1} 
  \partial^2_\mu \partial^2_{\rho} U  U^{-1}.
\end{equation}

Diagram 5.b:

Its divergent parts are
\begin{equation}
 \frac18 \theta(A - B)\left\{\Delta^2 (0) (3+z^2) 
+ \Delta(A,B) \Delta (0) (1-z^2)\right\} U^{-1} 
  \partial^2_\mu \partial_{\rho} U  U^{-1} \partial_{\rho} U  U^{-1}.
\end{equation}

Diagram 5.c:

Its divergent parts are
\begin{equation}
 \frac18 \theta(A - B) \left\{ \Delta^2 (0) (3+z^2) 
+ \Delta(A,B) \Delta (0) (1-z^2)
\right\} U^{-1} 
 \partial_{\rho} U  U^{-1}  \partial^2_\mu \partial_{\rho} U  U^{-1}.
\end{equation}

Diagram 5.d:

Its divergent parts are
\begin{eqnarray}
&& -\frac18 \theta(A - B) \left\{ \Delta^2 (0) (3+z^2) 
+ \Delta(A,B) \Delta (0) (1-z^2)\right\}\nonumber\\
&&\times U^{-1} 
 \partial_{\rho} U  U^{-1}  \partial^2_\mu U  U^{-1} \partial_{\rho} U  U^{-1}.
\end{eqnarray}

Diagram 5.e:

Its divergent and finite parts are
\begin{eqnarray}
&&-  \frac{1}{16} \theta(A - B)\left\{ 
\Delta^2 (0) (11 +z +5z^2 -z^3) 
+\Delta(A,B) \Delta (0)2 (1-z^2) (1-z)
\right.\nonumber\\
&&\left. + \Delta^2 (A,B)  (1 - z^2) (3+z)\right\}
U^{-1} \partial_{\rho} U  U^{-1}
 \partial_{\mu} U  U^{-1}  \partial_\mu \partial_{\rho} U  U^{-1}.
\end{eqnarray}

Diagram 5.f:

Its divergent and finite parts are
\begin{eqnarray}
&&-  \frac{1}{16} \theta(A - B)\left\{ 
\Delta^2 (0)  (11 -z +5z^2 +z^3)
+\Delta(A,B) \Delta (0)2 (1-z^2) (1+z)
\right.\nonumber\\
&&\left. + \Delta^2 (A,B)  (1 - z^2) (3-z)\right\}
U^{-1} \partial_{\mu}\partial_{\rho} U  U^{-1}
 \partial_{\mu} U  U^{-1} \partial_{\rho} U  U^{-1}.
\end{eqnarray}

Diagram 5.g:

Its divergent and finite parts are
\begin{equation}
\frac{1}{8} \theta(A - B)\left\{ \Delta^2 (0)(3+z^2) 
+  \Delta^2 (A,B) (1-z^2) \right\} U^{-1} 
 \partial_{\rho} \partial_\mu U  U^{-1} \partial_{\rho} \partial_\mu U  U^{-1}.
\end{equation}

Diagram 5.h:

Its divergent and finite parts are
\begin{eqnarray}
&& \frac{1}{32} \theta(A - B) \left\{\Delta^2 (0) (3 + z^2)(7 + z^2) 
 + 2(1-z^2)(3+z^2)\Delta (A,B)\Delta (0)\right.\nonumber\\ 
&&\left. + (1-z^2)(5-z^2)\Delta^2 (A,B)\right.\nonumber\\ 
&&\left. + 2 (1-z^2)^2\int^A_B d\tau_1 \int^{\tau_1}_B d\tau_2   
\partial_{\tau_1}\Delta (\tau_1,\tau_2) \partial_{\tau_2}\Delta (\tau_1,\tau_2)\right\}\nonumber\\
&&\times U^{-1} 
 \partial_{\rho} U  U^{-1} \partial_\mu U  U^{-1} \partial_\mu U  U^{-1}\partial_{\rho} U  U^{-1}.
\label{2-0,8}
\end{eqnarray}

Diagram 5.i

Its divergent and finite parts are
\begin{eqnarray}
&& -  \frac12 \theta(A - B)\left[ \Delta^2 (0) (1 + z^2) +\Delta (0) \Delta (A,B)(1-z^2)\right]\nonumber\\
&& \times U^{-1} 
 \partial_{\mu} U  U^{-1} \partial_\mu \partial_\rho U  U^{-1}\partial_{\rho} U  U^{-1}.
\end{eqnarray}

Diagram 5.j:

Its divergent and finite parts are
\begin{eqnarray}
&& \frac{1}{16} \theta(A - B) \left\{\Delta^2 (0) (1 + z^2)(9 - z^2) 
 + 2(1-z^2)(3-z^2)\Delta (A,B)\Delta (0)\right.\nonumber\\ 
&&\left. + (1-z^4)\Delta^2 (A,B) - (1-z^2)^2 \int^A_B d\tau \left(\Delta (A,\tau) - \Delta (A,B)\right) \dot{\Delta} (\tau,B)\right.\nonumber\\ 
&&\left. - (1-z^2)^2\int^A_B d\tau_1 \int^{\tau_1}_B d\tau_2   
\partial_{\tau_1}\Delta (\tau_1,\tau_2) \partial_{\tau_2}\Delta (\tau_1,\tau_2)\right\}\nonumber\\
&&\times U^{-1} 
 \partial_{\rho} U  U^{-1} \partial_\mu U  U^{-1} \partial_\rho U  U^{-1}\partial_{\mu} U  U^{-1}.
\label{2-0,10}
\end{eqnarray}
The first integral is assembled into a non-singular expression.
The second integral  is prepared to have a weaker singularity than 
$\Delta^2(0)$ and  
it reveals a  singularity of $\Delta (0)$-type.

\section{} 
Let us first understand the singularities in  $J(A,B)$ with the help of 
Euclidean 2-dim scalar field propagator (\ref{reg})
in the cutoff regularization,
\begin{equation}
\Delta(\tau,\sigma) = - \alpha'\ln\left[(\tau^2 + \sigma^2 + R^2)\mu^2\right],
\qquad 
R \rightarrow 0.
\end{equation}
The regularization smears 
the $\log$ singularity, the normalization is taken to 
provide \\$- \partial^2 \Delta(\tau,\sigma) =
4\pi\alpha' \delta(\tau)\delta(\sigma) $ for $R \rightarrow 0$.
Thus on the boundary,
\begin{equation}
\Delta(\tau,\sigma= 0) = -\alpha'\ln\left[(\tau^2 + R^2)\mu^2\right],\qquad 
\partial_\sigma \Delta(\tau,\sigma= 0) = 0,
\end{equation}
the latter being in accordance with 
Neumann boundary conditions for open strings.
Evidently, the relation between divergences in DR and the present
 regularization is:
\begin{equation}
\Delta(0) \simeq - \alpha'\ln\left[R^2\mu^2\right] \leftrightarrow - \frac{2\alpha'}{\epsilon}.
\end{equation}
Of course this propagator is not necessarily exact in its finite part.
But the divergences extracted with its help should be universal.

With this ansatz it is easy to find that
\begin{equation}
J(A - B) = \alpha'^2\left\{- \frac{\pi(A-B)}{R}\ 
- 2\ln\left(R^2\mu^2\right)\right\} + \mbox{regular terms}.
\end{equation}
The first divergence is power-like and we neglect it  
(it should not appear in DR calculation). The second term is 
logarithmically divergent $ \simeq 2\alpha' \Delta(0)$ 
and we retain only this one.

Now let us perform the same calculation in the Dimensional Regularization.
First we define the integral (\ref{Jab}) in $2 + \epsilon$ dimensions. Evidently
we can do it consistently for the second expression in  (\ref{Jab}). Namely, we 
replace $\tau \rightarrow |\vec\tau| \equiv t$ with $\vec\tau$ being a $1 + \epsilon$
dimensional vector and integrate over the sphere   $|\vec\tau| \leq (A - B)$,
\begin{equation}
J_\epsilon (A-B)= - \frac12 \int_{|\vec\tau| \leq (A-B)} d^{1+\epsilon}\tau
\mu^{\epsilon} \, (A-B -|\vec\tau|) 
\left[\dot\Delta(|\vec\tau|)\right]^2. \label{Jabeps}
\end{equation}
Next we insert in (\ref{Jabeps}) the
derivative of the  string propagator (\ref{reg}) in $2 + \epsilon$ dimensions,
\begin{equation}
\Delta'(t) = - \alpha'
\epsilon\Gamma\left(\frac{\epsilon}{2}\right) 
\left(\frac{\mu\sqrt{\pi}}{\varphi}\right)^{-\epsilon}
t^{-1 - \epsilon},
\end{equation}
which leads to
\begin{eqnarray}
J_\epsilon (A-B)=&& - \frac12  (\alpha')^2
\left(\epsilon\Gamma\left(\frac{\epsilon}{2}\right)\right)^2 \Omega_{1+\epsilon} 
\left(\frac{\mu\sqrt{\pi}}{\varphi}\right)^{-2\epsilon}\nonumber\\
&& \times \int^{A-B}_0 dt (t\mu)^\epsilon (A-B-t)t^{-2 - \epsilon},
\end{eqnarray} 
where the angular volume,
\begin{equation}
\Omega_{1+\epsilon}= \frac{2 \pi^{\frac{1+\epsilon}{2}}}{\Gamma\left(\frac{1+\epsilon}{2}\right)}. 
\end{equation}
After integration one finds the following expression
\begin{eqnarray}
J_\epsilon (A-B)&=& -   (\alpha'\varphi^{-\epsilon})^2
\left(\epsilon\Gamma\left(\frac{\epsilon}{2}\right)\right)^2 
\frac{\sqrt{\pi}}{\Gamma\left(\frac{1+\epsilon}{2}\right)} 
\frac{\left((A-B)\mu\sqrt{\pi}\right)^{-\epsilon}}{\epsilon(1+\epsilon)}\nonumber\\
&\stackrel{\epsilon \to 0}{=}&  -
\frac{4(\alpha'\varphi^{-\epsilon})^2}{\epsilon} = 2
\alpha'\varphi^{-\epsilon} \Delta(0).
\end{eqnarray} 
Thus we have reproduced the same value of the constant $c_V$.

\section{}
 Let us take the most general set of operators which can appear
in the equations of motion (E.o.M.) with arbitrary constants. 
The equations of motion
 of dimension two (\ref{resym}) are assumed to hold and 
therefore we do not include any vertices containing the D'Alambertian
$\partial_\mu^2$. We would like to find out a set of constants which
supports the unitarity relation (\ref{unit}),\quad $U\delta U^\dagger\, = 
\, - \delta U U^\dagger$. The results are presented in the Table 4. 

\begin{table}[h]
\tbl{Comparison between the 
coefficients of the different chiral field structures  in the equations
of motion, their Hermitean conjugates  and those ones 
derived from a local Lagrangian (\ref{dimen4})}
{\begin{tabular}{@{}cccc@{}} \toprule
{\bf CF structure}&E.o.M. &(E.o.M.)$^\dagger$ &$\chi$-lagr. \\ \colrule
$\mu--\rho--\mu\rho $&$a_1$ &$ - a_2$ & $-2(2K_1 + K_2)$ \\  
$\mu\rho--\mu--\rho$&$a_2$ &$ - a_1$&$ -2(2K_1 + K_2)$\\  
$\mu--\mu\rho--\rho $&$a_3$ &$- a_3$ &$-4K_2$\\  
$\mu--\rho--\rho--\mu$&$a_4$ &$ a_1 + a_2 + a_4$
 &$2[(1- z^2)K_1 + K_2]$\\  
$\mu--\mu--\rho--\rho $&$a_5$ &$ a_3 +  a_5 $ &$-2z^2 K_2$\\ 
$\mu--\rho--\mu--\rho$&$a_6$ & $ a_1 + a_2 + a_3 + a_6$&$4[K_1 + K_2] $\\
\botrule
\end{tabular}}
\end{table}

One can see that the unitarity of $\delta U^{(4)}$ is provided for
$z^2 = -1$  only if
\begin{eqnarray}
&&a_1 = a_2 = -(4 K_1 + 2 K_2),\quad a_3 = -4 K_2,\quad
a_4 = - \frac12 (a_1 + a_2), \nonumber\\
&&a_5 = - \frac12 a_3,\quad
a_6 = - \frac12 (a_1 + a_2 + a_3).
\end{eqnarray}
Thus unitarity is achieved when the operators in the equation of motion are
derived from the local Lagrangian (\ref{dim4}) and {\it vice versa}.

\section{}
In this appendix we shall explore a somewhat different approach
and see how the results are fully equivalent to those presented
in the text. We shall integrate out the `quarks' and derive
an effective action in terms of external sources.

To derive the effective action let us supplement the Lagrangian 
(\ref{lag}) with the external sources for fermion fields
\begin{equation}
\tilde L_f= L_f +\bar J_L\psi_R+\bar J_R\psi_L
+\bar\psi_L J_R+\bar\psi_R J_L. \label{lagext}
\end{equation}
As this Lagrangian is quadratic in fields the effective action, 
\begin{equation}
e^{iS_{eff}(x)}=\int d\bar\psi d\psi e^{i\int \tilde L_f dt}
\end{equation}
is supported by
the solutions of classical equations,
\begin{eqnarray}
\dot\psi_R+\frac{1}{2} (1+z)U^{-1}\dot U\psi_R &=& iU^{-1}J_R,\\
\dot{\bar\psi_R}+\frac{1}{2}(1+z^*)\bar\psi_R\dot U^\dagger{U^\dagger}^{-1} 
&=& -i\bar J_R{U^\dagger}^{-1},\\
\dot\psi_L+\frac{1}{2}(1-z^*){U^\dagger}^{-1}\dot
U^\dagger\psi_L &=& i{U^\dagger}^{-1} J_L,\\
\dot{\bar\psi_L}+\frac{1}{2} (1-z)\bar\psi_L
\dot U U^{-1} &=& -i \bar J_L U^{-1}.
\end{eqnarray}
The solutions read
\begin{eqnarray}
\psi_R(\tau) &=& \int_{-\infty}^{\tau} d\tau_1
\left[ \mbox{\bf T}\exp\left(-\int_{\tau_1}^\tau d\tau_2 
\frac{1}{2} (1+z)U^{-1}(\tau_2)\dot
U(\tau_2)\right)\right]\nonumber\\&&\times  i U^{-1}(\tau_1) J_R(\tau_1),\\
\psi_L(\tau) &=& \int_{-\infty}^{\tau} d\tau_1
\left[ \mbox{\bf T}\exp\left(-\int_{\tau_1}^\tau d\tau_2 \frac{1}{2} 
(1-z^*){U^\dagger}^{-1}
(\tau_2)\dot{U^\dagger}(\tau_2)\right)\right] \nonumber\\&&\times i 
{U^\dagger}^{-1}(\tau_1) J_L(\tau_1),
\end{eqnarray}
and their complex conjugated partners.

Therefore the effective action takes the simple form:
\begin{eqnarray}
&&\int_{-\infty}^{+\infty}d\tau\bar J_L(\tau)\int_{-
\infty}^{\tau} d\tau_1
\left[\mbox{\bf T}\exp\left(-\int_{\tau_1}^\tau d\tau_2 
\frac{1}{2} (1+z)U^{-1}(\tau_2)\dot
U(\tau_2)\right)\right] \nonumber\\&&\times i U^{-1}(\tau_1) J_R(\tau_1)
\end{eqnarray}
In turn the full fermion propagator takes the form
\begin{eqnarray}
\langle \psi_R (\tau_1) \bar\psi_L(\tau_2)\rangle &=& 
U^{-1} [x_\mu(\tau_1)] \mbox{\bf T}\exp\left[
\frac12 (1-z)\int^{\tau_1}_{\tau_2}d\tau \dot{U} U^{-1}[x_\mu(\tau)]
 \right]\theta(\tau_1 - \tau_2),\nonumber\\
 &=& 
\mbox{\bf T}\exp\left[
- \frac12 (1+z)\int^{\tau_1}_{\tau_2} d\tau U^{-1} \dot{U}[x_\mu(\tau)]
 \right] \nonumber\\&&\times U^{-1} [x_\mu(\tau_2)]\theta(\tau_1 - \tau_2). \label{propa}
\end{eqnarray}
We stress that two pieces with T-exponentials 
in the last equality are identical.

Let us expand the first expression for the propagator in (\ref{propa})
\begin{eqnarray}
&&\langle \psi_R (\tau_1) \bar\psi_L(\tau_2)\rangle = \theta(\tau_1 - \tau_2)
U^{-1} [x_\mu(\tau_1)] \biggl[1 +
\frac12 (1-z)\int^{\tau_1}_{\tau_2} d\tau \dot{U} U^{-1}[x_\mu(\tau)]\nonumber\\
&&+ \frac14 (1-z)^2\int^{\tau_1}_{\tau_2} d\tau \int^{\tau}_{\tau_2} d\tau' \dot{U} U^{-1}[x_\mu(\tau)]
\dot{U} U^{-1}[x_\mu(\tau')] + \cdots\biggr],
\label{expans}
\end{eqnarray}
The second order in the expansion is sufficient to analyze one-loop divergences of the propagator. 
In turn we can develop the perturbative expansion around 
a background, $x(\tau) = x_0 + \tilde x(\tau) $,
\begin{eqnarray}
&&U^{-1} [x_\mu(\tau_1)] = U^{-1}(x_0)\biggl\{1 - 
\tilde x_\mu(\tau_1) (\partial_\mu U) U^{-1}(x_0)\nonumber\\ 
&&+ 
\frac12 \tilde x_\mu(\tau_1) \tilde x_\nu(\tau_1) 
[-(\partial_\mu\partial_\nu U) U^{-1}(x_0) +
2 (\partial_\mu U) U^{-1} (\partial_\nu U) U^{-1}(x_0)\biggr\}.
\label{exp}
\end{eqnarray}
Evidently the term with two derivatives on $U$ comes out only from
this part of expansion.
Further on one should also  evaluate:
\begin{eqnarray}
 \dot{U} U^{-1}[x_\mu(\tau)] &=& \dot{\tilde x}_\mu  \partial_\mu U  U^{-1} (x_0) + 
\dot{\tilde x}_\mu {\tilde x}_\nu\left[ \partial_\mu\partial_\nu U U^{-1} (x_0)\right. \nonumber\\
&&\left.- \partial_\mu U U^{-1} 
\partial_\nu U U^{-1} (x_0)\right] + \cdots
\end{eqnarray}
Now we insert the above two expansions into Eq.~(\ref{expans}) and retain only 
terms quadratic in
${\tilde x}_\mu$,
\begin{eqnarray}
&& U^{-1}(x_0)
\frac12  
[-\partial_\mu\partial_\nu U U^{-1}(x_0) +
2 \partial_\mu U U^{-1} \partial_\nu U U^{-1}(x_0)]\tilde x_\mu(\tau_1) \tilde x_\nu(\tau_1)\nonumber\\
&&-   \frac12 (1-z)  U^{-1} \partial_\mu U U^{-1} \partial_\nu U U^{-1}(x_0) 
\int^{\tau_1}_{\tau_2}d\tau\tilde x_\mu(\tau_1) \dot{\tilde x}_\nu(\tau)\nonumber\\
&&+ \frac12 (1-z)  U^{-1}[ \partial_\mu\partial_\nu U U^{-1} (x_0) - \partial_\mu U U^{-1} 
\partial_\nu U U^{-1} (x_0)]   
\int^{\tau_1}_{\tau_2}d\tau \dot{\tilde x}_\mu(\tau) \tilde x_\nu(\tau)\nonumber\\
&&+   \frac14 (1-z)^2  U^{-1} \partial_\mu U U^{-1} \partial_\nu U U^{-1}(x_0) 
\int^{\tau_1}_{\tau_2}d\tau \int^{\tau}_{\tau_2}d\tau' \dot{\tilde x}_\mu(\tau) \dot{\tilde x}_\nu(\tau')
\end{eqnarray}
After integration over $\tau, \tau'$ and averaging in $\tilde x_\mu(\tau)$ with the help
of formulas for the string propagator, having in mind that $\dot\Delta(0)= 0$ (i.e. the contribution
of the third line is equal zero) one obtains
the 1-loop part in the form written in the main text,
\begin{eqnarray}
&&\frac12 \theta(\tau_1 - \tau_2)U^{-1} \biggl[\Delta (0)  
\biggl\{- \partial^2_{\mu} U\nonumber\\
&&+ 
\frac{3 + z^2}{2}\partial_\mu U U^{-1}\partial_\mu U\biggr\}
+ \frac{(1-z^2)}{2} \Delta (\tau_1 - \tau_2)
\partial_\mu U U^{-1}\partial_\mu U \biggr] U^{-1}.
\end{eqnarray}

\end{document}